\documentclass[aps,pra,twocolumn,superscriptaddress,article,nofootinbib,showpacs,longbibliography]{revtex4-2}

\usepackage{amssymb}
\usepackage{amsmath,bm}
\usepackage{amssymb}
\usepackage{graphicx}
\usepackage{amsfonts}         
\usepackage{fancybox}   
\usepackage{csquotes}
\usepackage{dsfont}
\usepackage{physics}

\usepackage{caption}
\usepackage[usenames,dvipsnames]{xcolor}%http://en.wikibooks.org/wiki/LaTeX/Colors

\usepackage[pagebackref,  % this puts links to the page numbers where refs appear
bookmarks={false}, pdfauthor={John McGreevy}, pdftitle={Yay, physics!}]{hyperref}
\hypersetup{colorlinks=true,
linkcolor=BrickRed,
citecolor=Violet,
filecolor=OliveGreen,
urlcolor=RoyalBlue,
filebordercolor={.8 .8 1},
urlbordercolor={.8 .8 0}}%http://en.wikibooks.org/wiki/LaTeX/Hyperlinks
\usepackage{soul}%strike through text \st{} .  not in math mode.
\setstcolor{Red}
\definecolor{darkgreen}{rgb}{0,0.4,0}
\definecolor{darkred}{rgb}{0.4,0,0}
\definecolor{darkblue}{rgb}{0,0,0.4}
\definecolor{lightblue}{rgb}{.6,.6,0.9}

\definecolor{uglybrown}{rgb}{0.8,  0.7,  0.5}

\definecolor{palatinatepurple}{rgb}{0.41, 0.16, 0.38}
\definecolor{celebrationcolor}{rgb}{0.75,  0.0,  0.9}

\definecolor{shadecolor}{rgb}{0.90,0.90,0.90}
\definecolor{DVcolor}{rgb}{0.95,  0.5,  0.2}
\definecolor{lightbluemuons}{rgb}{0.0,.65,1.0}

\definecolor{chartreuse}{rgb}{0.70, 1.00, 0.00}

\usepackage{tikz}
\usetikzlibrary{shapes}
\usetikzlibrary{trees}
\usetikzlibrary{snakes}
\usetikzlibrary{matrix,arrows} 				% For commutative diagram
\usetikzlibrary{positioning}				% For "above of=" commands
\usetikzlibrary{calc,through}				% For coordinates
\usetikzlibrary{decorations.pathreplacing}  % For curly braces
\usepackage[tikz]{bclogo} 					% For cute logo boxes
\usepackage{pgffor}							% For repeating patterns

\usetikzlibrary{decorations.markings}
\usetikzlibrary{intersections}				% For interesections

\usetikzlibrary{arrows,decorations.pathmorphing,backgrounds,positioning,fit,petri,automata,shadows,calendar,mindmap, graphs}
\usetikzlibrary{arrows.meta,bending}

\tikzset{
    vector/.style={decorate, decoration={snake}, draw},
    fermion/.style={postaction={decorate},
        decoration={markings,mark=at position .55 with {\arrow{>}}}},
    fermionbar/.style={draw, postaction={decorate},
        decoration={markings,mark=at position .55 with {\arrow{<}}}},
    fermionnoarrow/.style={},
    gluon/.style={decorate,
        decoration={coil,amplitude=4pt, segment length=5pt}},
    scalar/.style={dashed, postaction={decorate},
        decoration={markings,mark=at position .55 with {\arrow{>}}}},
    scalarbar/.style={dashed, postaction={decorate},
        decoration={markings,mark=at position .55 with {\arrow{<}}}},
    scalarnoarrow/.style={dashed,draw},
	vectorscalar/.style={loosely dotted,draw=black, postaction={decorate}},
}

\def\centerarc[#1](#2)(#3:#4:#5)% Syntax: [draw options] (center) (initial angle:final angle:radius)
    { \draw[#1] ($(#2)+({#5*cos(#3)},{#5*sin(#3)})$) arc (#3:#4:#5); }

\usepackage{enumitem}

\usepackage{slashed}

\usepackage[font=small,labelfont=bf]{caption}
\DeclareCaptionFont{tiny}{\tiny}
\usepackage{epstopdf}
\usepackage{wasysym}

\usepackage[yyyymmdd,hhmmss]{datetime}   
\usepackage{framed}  % needed for \begin{shaded}
 \usepackage{mdframed}
 \usepackage{wrapfig}
 \usepackage{yfonts}  
\newmdenv[%
        backgroundcolor=lightgray,
    linecolor=black,
    outerlinewidth=2pt,
]{boxedandshaded}

\numberwithin{equation}{section}

\newlength{\extraspace}
\newlength{\extraspaces}
\def\be{\begin{equation}}
\def\ee{\end{equation}}

\newcommand{\bea}{\begin{eqnarray}}
\newcommand{\eea}{\end{eqnarray}}

\def\Tr{{{\rm Tr~ }}}
\def\tr{{\rm tr}}
\def\Re{{\rm Re\hskip0.1em}}
\def\Im{{\rm Im\hskip0.1em}}

\def\bra#1{\left\langle#1\right|}
\def\ket#1{\left|#1\right\rangle}

\def\CG{{\cal G}}

\def\CO{{\cal O}}%AEL

\def\II{\relax{I\kern-.10em I}}

\def\IB{\relax{\rm I\kern-.18em B}}

\def\ID{\relax{\rm I\kern-.18em D}}
\def\IE{\relax{\rm I\kern-.18em E}}
\def\IF{\relax{\rm I\kern-.18em F}}
\def\IG{\relax\hbox{$\inbar\kern-.3em{\rm G}$}}
\def\IGa{\relax\hbox{${\rm I}\kern-.18em\Gamma$}}
\def\IH{\relax{\rm I\kern-.18em H}}
\def\II{\relax{\rm I\kern-.18em I}}
\def\IK{\relax{\rm I\kern-.18em K}}
\def\inbar{\,\vrule height1.5ex width.4pt depth0pt}

\def\IR{\mathbb{R}}

\def\lp10{\ell_p^{10}}
\def\lp11{\ell_p^{11}}
\def\R11{R_{11}}

\def\frac#1#2{{#1 \over #2}}

\newdimen\tableauside\tableauside=1.0ex
\newdimen\tableaurule\tableaurule=0.4pt
\newdimen\tableaustep
\def\phantomhrule#1{\hbox{\vbox to0pt{\hrule height\tableaurule width#1\vss}}}
\def\phantomvrule#1{\vbox{\hbox to0pt{\vrule width\tableaurule height#1\hss}}}
\def\sqr{\vbox{%
  \phantomhrule\tableaustep
  \hbox{\phantomvrule\tableaustep\kern\tableaustep\phantomvrule\tableaustep}%
  \hbox{\vbox{\phantomhrule\tableauside}\kern-\tableaurule}}}
\def\squares#1{\hbox{\count0=#1\noindent\loop\sqr
  \advance\count0 by-1 \ifnum\count0>0\repeat}}
\def\tableau#1{\vcenter{\offinterlineskip
  \tableaustep=\tableauside\advance\tableaustep by-\tableaurule
  \kern\normallineskip\hbox
    {\kern\normallineskip\vbox
      {\gettableau#1 0 }%
     \kern\normallineskip\kern\tableaurule}%
  \kern\normallineskip\kern\tableaurule}}
\def\gettableau#1 {\ifnum#1=0\let\next=\null\else
  \squares{#1}\let\next=\gettableau\fi\next}

\def\({\left(}
\def\){\right)}

\def\ii{{\rm i}}

\def\lsim{\mathrel{\mathstrut\smash{\ooalign{\raise2.5pt\hbox{$<$}\cr\lower2.5pt\hbox{$\sim$}}}}}
\def\gsim{\mathrel{\mathstrut\smash{\ooalign{\raise2.5pt\hbox{$>$}\cr\lower2.5pt\hbox{$\sim$}}}}}

\def\overleftrightarrow#1{\vbox{\ialign{##\crcr
     $\leftrightarrow$\crcr\noalign{\kern-0pt\nointerlineskip}
     $\hfil\displaystyle{#1}\hfil$\crcr}}}
     
     \def\overleftarrow#1{\vbox{\ialign{##\crcr
     $\leftarrow$\crcr\noalign{\kern-0pt\nointerlineskip}
     $\hfil\displaystyle{#1}\hfil$\crcr}}}

\def\gO{\textsf{O}}
\def\gSU{\textsf{SU}}
\def\gU{\textsf{U}}

\newif{\ifeq}           % defines a new condition @eq tested by the conditional \ifeq
\eqtrue                 % if uncommented, declares @eq to be true
\newcounter{lecturecounter}
\usepackage{siunitx}
\usepackage{braket}
\usepackage{multirow}
\usepackage{makecell}

\newcommand{\eq}[1]{\begin{equation}#1\end{equation}}

\newcommand{\eqnref}[1]{Eq.\,\eqref{#1}}
\newcommand{\figref}[1]{Fig.\,\ref{#1}}
\newcommand{\tabref}[1]{Tab.\,\ref{#1}}
\newcommand{\secref}[1]{Sec.\,\ref{#1}}
\newcommand{\appref}[1]{App.\,\ref{#1}}

\newcommand{\vect}[1]{{\bm{#1}}}
\newcommand{\dsi}{\mathds{1}}
\newcommand{\bk}{\vect{k}}
\newcommand{\bK}{\vect{K}}
\newcommand{\ki}[1]{\vect{k}_{#1}}
\newcommand{\Ki}[1]{\vect{K}_{\tau_{#1}\alpha_{#1}}}
\newcommand{\momind}[1]{\tau_#1\alpha_#1}
\newcommand{\bd}{d}
\newcommand{\boldp}{\vect{p}}
\newcommand{\boldq}{\vect{q}}
\newcommand{\tJ}{\tilde{J}}
\newcommand{\tV}{\tilde{V}}

\newcommand{\LUV}{\Lambda_{\text{UV}}}

\newcommand{\ivct}{\mathcal{I}_t}
\newcommand{\ivcs}{\mathcal{I}_s}

\newcommand{\taua}[1]{\tau_{#1}\alpha_{#1}}

\hypersetup{
  pdfauthor = {Da-Chuan Lu},
  pdftitle = {Correlated metals and unconventional superconductivity in rhombohedral trilayer graphene: a renormalization group analysis}
}

\begin{document}
\title{Correlated metals and unconventional superconductivity in rhombohedral trilayer graphene: a renormalization group analysis}
\author{Da-Chuan Lu}
\affiliation{Department of Physics, University of California, San Diego, CA 92093, USA\looseness=-1}
\author{Taige Wang}
\affiliation{Department of Physics, University of California, Berkeley, CA 94720, USA}
\affiliation{Material Science Division, Lawrence Berkeley National Laboratory, Berkeley, CA 94720, USA}
\author{Shubhayu Chatterjee}
\affiliation{Department of Physics, University of California, Berkeley, CA 94720, USA}
\author{Yi-Zhuang You}
\affiliation{Department of Physics, University of California, San Diego, CA 92093, USA\looseness=-1}

\begin{abstract}
Motivated by recent experimental observations of correlated metallic phases and superconductivity in rhombohedral trilayer graphene (RTG), we perform an unbiased study of electronic ordering instabilities in hole-doped RTG. Specifically, we focus on electronic states energetically proximate to Van Hove singularities (VHSs), where a large density of states promotes different interaction-induced symmetry-breaking electronic orders. To resolve the Fermi surface near VHSs, we construct a fermionic hot-spot model and demonstrate that a perpendicular electric field can tune different nesting structures of the Fermi surface. Subsequently, we apply a renormalization group analysis to describe the low-energy phase diagrams of our model under both short-range repulsive interactions as well as realistic (long-range) Coulomb interactions. Our analysis shows instabilities towards either intervalley coherent metallic phases or superconducting phases. The dominant pairing channel depends crucially on the nature of Fermi surface nesting --- repulsive Coulomb interaction favors spin-singlet $d$-wave pairing for relatively small displacement field and spin-singlet $i$-wave pairing for larger displacement field. We argue that the phase diagram of RTG can be well-understood by modeling the realistic Coulomb interaction as the sum of repulsive density-density interaction and ferromagnetic spin-triplet intervalley coherence (IVC) Hund's coupling, while phonon-mediated electronic interactions have a negligible effect on this system, in sharp contrast to twisted graphene multilayers.
% changed ud = ... to relatively small / larger displacement field
\end{abstract}
%\pacs{Keywords: }
\maketitle

\section{Introduction}

The recent observations of superconductivity and spin-valley symmetry broken phases in rhombohedral trilayer graphene (RTG) \cite{trg_exp2021half,trg_exp2021superconductivity,AndreaBilayer} establish it as an interesting platform for studying strong electronic correlations in graphene-based materials \cite{feldman2009gph1,mayorov2011gph2}. Previous studies on twisted multilayer graphene have found a rich phase diagram, including robust superconductivity, arising from enhanced electronic interactions due to the large density of states (DOS) in the flat bands hosted by these materials
\cite{cao2018moire1,cao2018moire2,yankowitz2019moire3,lu2019moire5,chen2019moire6,liu2020moire7,arora2020moire8,andrei2020moire9,balents2020moire10,hao2021moire11,park2021moire12}.
%Want to avoid repeating strong correlations in two back to back sentences
Compared to such moir\'e graphene systems, RTG samples are remarkably clean and free from local strains;
%more uniform and with less strain,
thus RTG is much more accessible and reproducible experimentally \cite{CaltechRTG}, while still featuring a high electronic DOS
%at low doping level
under a large vertical electric field (also called a displacement field).
More specifically, the band structure of RTG contains Van Hove singularities (VHSs) with divergent DOS in the vicinity of charge neutrality \cite{zhang2010bandtrg1}, near which several isospin (spin/valley) polarized metallic phases have been found via quantum oscillation measurements \cite{trg_exp2021half,trg_exp2021superconductivity}. There are two superconducting phases (SC) at the phase boundary of two distinct isospin symmetry-broken phases. In particular, the large region of the so-called SC1 phase appears to be a spin-singlet SC closely proximate to a spin-unpolarized symmetry-breaking metallic phase \cite{trg_exp2021superconductivity}.

Among many theoretical proposals \cite{ivcchatterjee2021rtg,SarmaTrilayer,SarmaBilayer,LevitovRTG,ErezRTG,ivcyizhuangyou2021KLsuperconductivityinRTG,MacDonaldRTG,qin2022frg_trg,szabo2022metalsrtg,TianxingRTG}, one promising pairing mechanism of the superconductivity in multilayer graphene systems is the Kohn-Luttinger mechanism that drives the initial repulsive electron interaction to change sign \cite{kohn1965KLsc,Chubukov1993}. Using the renormalization group (RG) approach, prior study in Ref.~\onlinecite{ivcyizhuangyou2021KLsuperconductivityinRTG} identified spin-singlet and spin-triplet superconductivity in RTG in different parameter regimes. However, the pairing symmetry of the SC phase remains unclear since the RG approach in Ref.~\onlinecite{ivcyizhuangyou2021KLsuperconductivityinRTG} lacked further resolution of the Fermi surface within each valley, e.g., $s,d$ or $p,f$ wave pairing could not be distinguished. Moreover, the consequences of two distinct forms of symmetry-allowed intervalley Hund's coupling, namely,
the spin Hund's coupling \cite{ivcyizhuangyou2021KLsuperconductivityinRTG} and the inter valley coherence (IVC) Hund's coupling \cite{ivcchatterjee2021rtg} proposed in the previous works are different and yet to be determined beyond perturbative considerations. In spite of their similar appearance, they can have a drastically different impact on the physical phase diagram, e.g., by favoring different pairing symmetries.

In this work, we use the hot-spot renormalization group developed in Ref.~\onlinecite{isobe2018unconventional} as well as other subsequent works \cite{lin2019chiraltbg,lin2020parquettbg} to study the infrared phases of hole-doped RTG at doping close to the VHSs under different types of electron interactions.
While previous studies often assumed a hierarchy of interaction scales, our RG approach enables us to treat all symmetry-allowed interaction vertices on an equal footing.
% We consider three different nesting structures in the experimental parameter regime, and develop a systematic approach to generate RG equations that respect the
% lattice symmetry and spin/valley symmetry.
Within our hot-spot model, we find that the perpendicular electric field can tune between three distinct nesting structures of the Fermi surfaces.
For each scenario, we develop a systematic approach to generate RG equations that respect the space group (lattice) and isospin (spin/valley) symmetries.
We find that ferromagnetic (antiferromagnetic) spin-Hund's coupling favors spin-triplet $f$-wave (spin-singlet $s$-wave) pairing, while ferromagnetic spin-triplet IVC Hund's coupling favors spin-singlet $d/i$-wave pairing depending on the nesting structure.
The ferromagnetic spin-singlet IVC Hund's coupling favors either spin-singlet IVC ordering or spin-triplet $p,f'$-wave pairing depending on the nesting structure.

% Need to connect to the previous paragraph
General understanding of electronic instabilities from RG flows in hand, we consider the physically relevant interaction vertices in RTG.
To this end, we project the realistic Coulomb interaction onto the interaction vertex basis.
We find that it is a combination of a repulsive density-density interaction and a small ferromagnetic spin-triplet IVC Hund's coupling, which favors spin-singlet $d/i$-wave pairing for different nesting structures.
We also consider phonon-mediated electronic interaction, which takes the form of an attractive density-density term.
However, in RTG it is parametrically quite weak compared to the Coulomb interaction.
Therefore, when we take into account both Coulomb repulsion and phonon-mediated attraction
%adding the phonon mediated interaction and Coulomb interaction
as bare interactions in the renormalization group analysis, the weak phonon-mediated interaction gets dominated by the repulsive density-density interaction part of the strong Coulomb term, and the effect of the phonon-mediated interaction is negligible. Consequently, the infrared fate of RTG is always characterized by a spin-singlet $d/i$-wave superconductor.

The remainder of this paper is organized as follows. In \secref{sec:band_hotspot}, we present the Hamiltonian that determines the band structure of RTG and then projects the Hamiltonian to the Fermi pockets in the vicinity of VHSs. We discuss the nesting structures of these Fermi surfaces and the various resulting instabilities in \secref{sec:nesting}. In \secref{sec:intvert}, we present a systematic way to represent the four fermion interactions with respect to lattice symmetry and spin/valley symmetry by constructing the interaction vertex basis in \secref{sec:int_basis}. We discuss different types of local interactions, especially the spin Hund's and IVC Hund's coupling in \secref{sec:interactions} and realistic Coulomb and phonon mediated electron interaction in \secref{sec:coulomb}. The hot-spot RG is introduced in \secref{sec:RG}, and the procedure to generate RG equations is presented in \secref{sec:RGE}. We present the infrared phase diagrams under these interactions and discuss the effect of each interaction term in \secref{sec:pd}. Finally, we conclude in \secref{sec:concl} with a summary of our main results and an outlook for future experiments.

\section{Band structure and hot-spot model}\label{sec:band_hotspot}

\begin{figure}[htbp]
    \centering
    \includegraphics[width=0.45\textwidth]{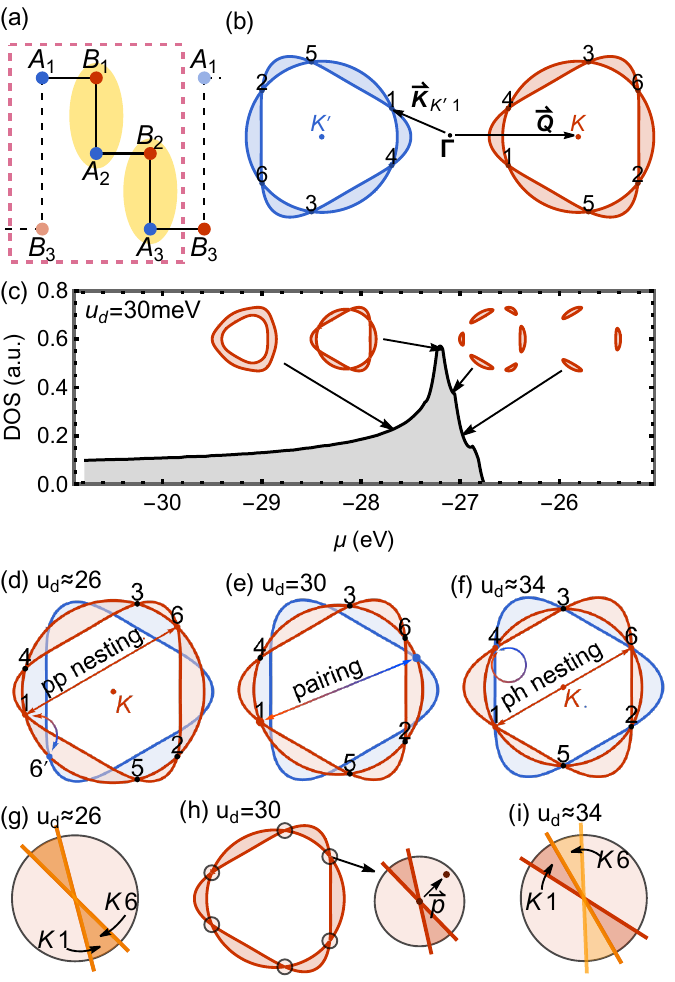}
    \caption{(a) shows the unit-cell of RTG, where the $B_1/A_2$ and $B_2/A_3$ are strongly hybridized such that the active sublattices $A_1/B_3$ form a triangular lattice. (b) shows the Fermi surfaces around $K$ (red) and $K'$ (blue) valley, the momentum $\Ki{},\vect{Q}$ are illustrated schematically. (c) Density of state (DOS) v.s. the chemical potential in the two band model with displacement field $u_d=\SI{30}{\meV}$. Insets show the $K$ valley Fermi surfaces at the corresponding doping level. The annular Fermi surface transitions to Fermi pockets via the Van Hove singularity. (d), (e), (f) show the Fermi surfaces with VHSs under different displacement fields $u_d$. The generic Fermi surface (e) at $u_d=\SI{30}{\meV}$ only has Cooper pairing. The Fermi surface (d) at $u_d\approx \SI{26}{\meV}$ has additional particle-particle nesting as shown in (g), where the Fermi surfaces near VHS 1, 6 in the $K$ valley (or $K1$ and $K'6$) overlap under the translation. (f) at $u_d\approx \SI{34}{\meV}$ has additional particle-hole nesting as shown in (i), where the Fermi surfaces near $K1,K6$ (or $K1,K'6$) share the same edge. (h) shows the patches around the hot-spots which are used to implement the renormalization group calculation, the momentum relative to the hot-spot is $\vect{p}$.}
    \label{fig:fsandnesting}
\end{figure}

Rhombohedral trilayer graphene (RTG) consists of three layers of graphene with ABC-stacking. Its electronic band structure is accurately modeled by six-band model per valley and spin \cite{zhang2010bandtrg1,jung2013bandtrg2,ho2016bandtrg3}. However, the low-energy electronic states that are the most relevant to the various ordering instabilities can be described well by a relatively simpler two-band model. In the following calculations, we focus on the two-band model, which captures essential low-energy features of the more complicated six-band model. The electronic states near Fermi level mainly reside on top $A_1$ and bottom $B_3$ sites, the effective two-band model is, $H_0=\sum_{\tau,\bk,s,\sigma} c_{\tau\bk s\sigma}^\dagger [h_{\tau}(\bk)]_{\sigma,\sigma'}c_{\tau\bk s\sigma'}$, where $c^\dagger,c$ are electron creation and annihilation operators, $\tau = K/K'$ labels valley, $s=\uparrow/\downarrow$ labels spin and $\sigma=A_1/B_3$ is the sublattice index and related to the top/bottom layer as shown in \figref{fig:fsandnesting} (a). The full form of the two-band Hamiltonian can be found in \appref{app:fullham}. The Hamiltonian contains two model parameters: the displacement field $u_d$ (the perpendicular electric field) and the total chemical potential $\mu$, which are experimentally tunable by applying different gate voltages in a dual gate setup.

At a fixed displacement field $u_d$, upon hole doping, the spin-degenerate Fermi surface in each valley transitions from three (or six) pockets related by $C_3$ rotation to an annular Fermi surface through a Van Hove singularity (VHS) where the density of states (DOS) diverges logarithmically, as shown in \figref{fig:fsandnesting} (c). The most significant contributions to the diverging DOS come from the electronic states around the band dispersion saddle points in the momentum space where the Fermi pockets touch and merge together to form the annular Fermi sea. These saddle points will also be referred to as hot-spots or ordinary Van Hove singularity points (VHSs) \cite{yuan2019vhs}, where the adjective ``ordinary'' is to indicate that they are not higher-order VHSs \cite{yuan2019vhs}. In RTG, there are typically six hot-spots per valley.

%why vhs
Various low-energy instabilities, such as superconductivity and spin-valley symmetry breaking order, can arise from the almost-degenerate electronic states around these hot-spots. Indeed, experiments found that the superconductivity occurs proximate to the isospin breaking phase on the hole-doped side, where the chemical potential rests in the regime between the annular Fermi surface and the disjoint Fermi pockets, i.e., near the VHSs. To investigate the instabilities caused by these hot-spots, for any given displacement field $u_d$, we fine-tune the chemical potential $\mu$ to the energy level of VHS, and perform a hot-spot renormalization group (RG) analysis.

\subsection{Hot-Spot Model}
We denote the momentum space location of the $\alpha$th VHS in valley $\tau$ by $\bK_{\tau\alpha}$ ($\alpha=1,2,\cdots,6$, $\tau=K,K'$). The Brillouin zone momentum $\vect{k}$ can then be decomposed as $\bk=\bK_{\tau\alpha}+\boldp$, where $\boldp$ is the momentum relative to the VHS. The low-energy effective Hamiltonian can be obtained by projecting the two-band Hamiltonian onto the valence band and expanding around these VHSs,
\begin{equation}\label{eq:hotspotham}
    H_0=\sum_{\tau,\alpha,\boldp,s} \psi_{\tau\alpha\boldp s}^\dagger \epsilon^{\tau\alpha}_\boldp \psi_{\tau\alpha\boldp s},
\end{equation}
where $\epsilon^{\tau\alpha}_\boldp$ is the energy dispersion near the VHS labeled by the valley index $\tau$ and the hot-spot index $\alpha$, and $\psi_{\tau\alpha\boldp s}$ are the Bloch electron creation and annihilation operator such that $c_{\tau \bk s \sigma}^{\dagger}=u_{\tau \bk s}^{*}(\sigma) \psi_{\tau \bk s}^{\dagger}$, with $u_{\tau \bk s}^{*}(\sigma)$ being the Bloch wave-function of the valance band which is relevant to hole-doped RTG.
%and $\boldp$ is the momentum relative to the VHSs $\bK_{\tau\alpha}$ in the momentum space.
To the leading order of $\vect{p}$, the energy dispersion $\epsilon^{\tau\alpha}_\boldp=(\vect{t}^{\tau\alpha}_{+}\cdot\vect{p})(\vect{t}^{\tau\alpha}_{-}\cdot\vect{p})$ takes the general form of a hyperbolic surface, where $\vect{t}^{\tau\alpha}_{\pm}$ specifies the tangent directions of the two crossing Fermi surfaces near the hot-spot.

\subsection{Nesting Structures of the Fermi surface}\label{sec:nesting}

The Fermi surface nesting structure crucially affects the type of leading instability. Here we investigate three representative nesting structures of the Fermi surface at the VHS in the valence band of the RTG system, which correspond to $u_d\approx 26,30,\SI{34}{\meV}$ respectively. Since the contribution to low energy physics is dominated by electrons near the VHSs, the nesting we considered here only concerns the Fermi surfaces very close to the VHSs, where linearized dispersion is a good approximation for ordinary VHSs.

As shown in \figref{fig:fsandnesting} (e), the generic Fermi surface within the experimental parameter regime can be represented by $u_d\approx \SI{30}{\meV}$, which contains six hot-spots per valley. The hot-spots in opposite valleys are related by time-reversal symmetry. The generic case does not have any nesting, therefore, the only diverging channel is the Cooper pairing channel.

There are two special cases, when $u_d\approx \SI{26}{\meV}$, \figref{fig:fsandnesting} (d) (g) shows Fermi surfaces around hot-spots 1, 6 and other symmetry related pairs overlap with each other, this causes additional diverging susceptibility in the particle-particle channel with momentum transfer that connects $K1,K6$ or $K1,K'6$. The effect of this particular nesting is different from the Cooper pairing since this type of particle-particle nesting connects more pairs of VHSs, therefore, it opens the possibility to have superconductivity with less symmetry. We will show in some parameter regimes of the fermion interactions, that the infrared phases would be $p$-wave pairing and $d$-wave pairing.

Additionally, \figref{fig:fsandnesting}(f) (i) shows the six hot-spots are related by an emergent $C_6$ rotation symmetry when $u_d\approx \SI{34}{\meV}$, the Fermi surface around 1, 6 in $K$ valley and other symmetry related pairs share the common edge, which causes diverging susceptibility in the particle-hole channel and favors charge ordering. Indeed, in some parameter regimes, the IR phase appears to be the spin-singlet intervalley coherent state (IVC), which microscopically corresponds to a charge density wave.

\section{Interaction vertices}\label{sec:intvert}
The electrons in the system can interact with each other via the Coulomb interaction or other interactions mediated by fluctuating bosonic fields, such as phonons. Those interaction terms can be generically written as four-fermion interaction. As discussed in the previous section, only the electrons near the VHS will contribute most to low-energy physics. We assume the general interaction term is projected to the closet hot-spots. Therefore, the interaction vertices do not have continuous momentum dependence. Due to momentum conservation, four fermion interaction vertex will be generally labeled by three hot-spot momenta, as $V(\Ki{1},\Ki{2},\Ki{3})$, where $\Ki{}$ denotes the momentum of the hot-spot $\alpha$ in $\tau$ valley.

\subsection{Interaction Vertex basis}\label{sec:int_basis}
The interaction vertices in the Hamiltonian should obey the global symmetries of the system. Besides the momentum conservation, the system also have reflection symmetry with respect to $x$-axis $R_x (k_x,k_y)=(k_x,-k_y)$, time reversal symmetry $T (k_x,k_y)=(-k_x,-k_y)$, and three-fold rotation symmetry $C_3$ which can be enlarged to $C_6$ in the special case of $u_d\approx \SI{34}{\meV}$. These discrete symmetries relate to different hot-spots. Then the interaction vertices should be identified if they are related by symmetries, $V(\Ki{1},\Ki{2},\Ki{3}) = V(g\Ki{1},g\Ki{2},g\Ki{3}),\quad \forall g \in G$, where $G=T\times R_x\times C_{3/6}$. We omit the fourth momentum since it is fixed by momentum conservation. Under this identification, the non-equivalent interaction vertices can be labeled by representatives among the momentum conserved interaction vertices, and the group action of $G$ on these representatives forms the orbits of the representatives in the group theory sense.

The low-energy Hamiltonian also possesses valley $\gU(1)$ particle number conservation symmetry, and spin $\gSU(2)$ symmetry which can be broken down to spin $\gU(1)$, for example, in presence of easy-axis or easy-plane anisotropy. When adding back the spin indices to interaction vertices, the interaction vertices obeying the spin $\gU(1)$ and valley $\gU(1)$ symmetry fall into 9 categories, as labeled by  $\vect{\eta}^s,\vect{\eta}^v=\{(1,1),(1,-1),(-1,1)\}$,
\begin{equation}\label{eq:intverts}
    \begin{gathered}
    \includegraphics[width=0.2\textwidth]{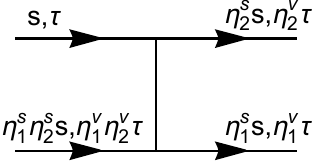}
    \end{gathered}
\end{equation}
where $-\tau,-s$ refer to the opposite valley or spin. We systematically enumerate all the interaction vertices obeying the lattice symmetry $G$, spin $\gU(1)$, and valley $\gU(1)$ symmetries. For six hot-spots per valley, there are 99 inequivalent interaction vertices, and they span the space of interaction vertices. However, they are not linearly independent, for example, $\psi_4^\dagger \psi_3^\dagger \psi_2 \psi_1$ is equal to $-\psi_3^\dagger \psi_4^\dagger \psi_2 \psi_1$. By Gaussian elimination, we obtain 43 linearly independent basis elements, the details can be found in \appref{app:intervert}. These 43 linearly independent basis elements span the space of momentum conserved interaction vertices with respect to discrete symmetries and at least spin $\gU(1)$, valley $\gU(1)$ symmetry. The set of interaction vertices in presence of enhanced symmetry, e.g., spin $\gSU(2)$ symmetry, then corresponds to a special subspace in this space of interaction vertices. In the RG calculation, we will keep track of the flow of these 43 interaction vertices among the hot-spots.

The linearly independent interaction vertex basis can be grouped into 5 categories as shown in \figref{fig:interactions}. The first two categories are intravalley processes, where the first one also keeps the spin indices. Due to the anti-symmetry of the interaction vertices and the fixed valley, spin indices, the only possible process in the first category corresponds to a finite momentum process among four different hot-spots. The second category corresponds to local intravalley density-density interactions. The last three categories are intervalley scattering processes with and without spin-flips.

\subsection{Hund's coupling and other interactions}\label{sec:interactions}
Among the four-fermion interactions, the intervalley Hund's coupling \eqnref{eq:HJIVC} is essential for determining the isospin symmetry-breaking phase \cite{trg_exp2021half,trg_exp2021superconductivity}. Moreover, a ferromagnetic Hund's coupling is consistent with spin-polarized phases found in the phase diagram of RTG. Short-range density-density interactions, like an on-site Hubbard interaction, can give rise to intervalley Hund's coupling \cite{ivcchatterjee2021rtg}.

Before analyzing realistic Coulomb interaction and phonon-mediated interaction, we first build intuition from the local interaction whose strength is uniform in the momentum space, namely the coupling constant does not depend on the momentum transfer. The are two versions of Hund's couplings in the existing literature: the spin Hund's coupling
\begin{equation}\label{eq:HJspin}
    H_J = -J_H \sum_\boldq\vect{S}_K(\boldq)\cdot \vect{S}_{K'}(-\boldq),
\end{equation}
where $S_\tau^i(\boldq) = \sum_{\bk,ss'} \psi_{\tau\bk+\boldq s}^\dagger \sigma^i_{ss'} \psi_{\tau\bk s'}$; and the intervalley coherence (IVC) Hund's coupling
\begin{equation}\label{eq:HJIVC}
    H_{\tJ} = -\tJ_H \sum_\boldq \vect{I}(\boldq)^\dagger \cdot \vect{I}(\boldq),
\end{equation}
where $I^i(\boldq)=\sum_{\bk,ss'} \psi_{K'\bk+\boldq s}^\dagger \sigma^i_{ss'} \psi_{K\bk s'}$.
The IVC Hund's coupling can be derived from the short-range component of the Coulomb interaction \cite{ivcchatterjee2021rtg}. One can view this Hund's coupling as a projection of density-density interaction onto the valence band. Both forms of Hund's coupling break the independent spin rotation symmetries in each valley $\gSU(2)_K^s\times \gSU(2)_{K'}^s$ down to physical spin $\gSU(2)_s$ symmetry.

Besides the Hund's couplings, symmetry allowed interactions also include the intravalley density-density interaction
\begin{equation}
    H_U = U \sum_{\tau,\boldq} n_\tau(\boldq) n_\tau(-\boldq),
\end{equation}
where $n_\tau(\boldq)=\sum_{\bk s} \psi_{\tau\bk+\boldq s}^\dagger \psi_{\tau\bk s}$; and two forms of intervalley density-density interactions
\begin{align}
    &H_V = V \sum_\boldq n_{K}(\boldq)n_{K'}(-\boldq), \\
    &H_{\tV} = -\tV \sum_\boldq I^0(\boldq)^\dagger I^0(\boldq) ,
\end{align}
where $I^0(\boldq)=\sum_{\bk s} \psi_{K\bk+\boldq s}^\dagger \psi_{K'\bk s}$. $H_{\tV}$ describes the ferromagnetic spin-singlet IVC Hund's coupling when $\tV>0$.

\begin{figure*}[t]
\includegraphics[width=0.95\textwidth]{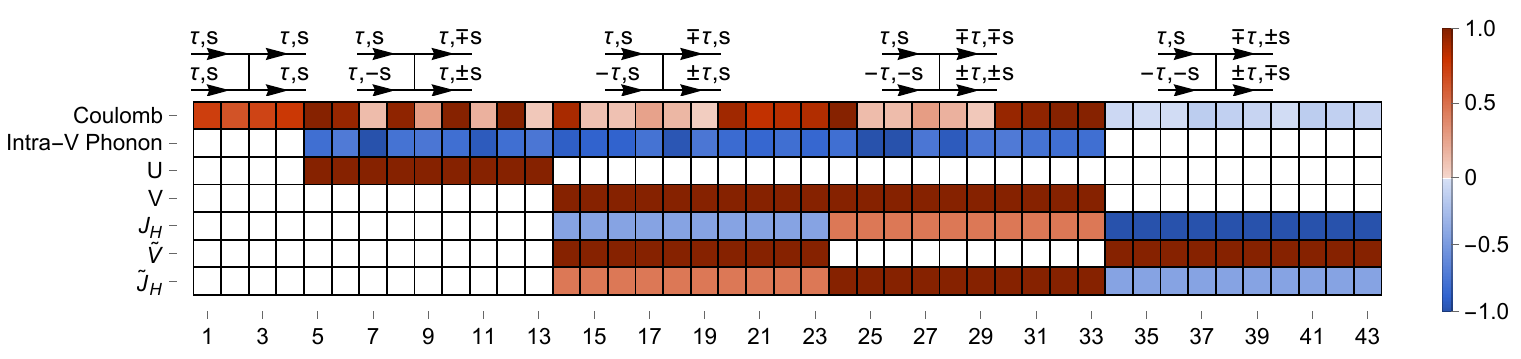}
\caption{Various interactions are projected onto interaction basis, the colored squares refer to the coefficients. The coefficients of each interaction vertex are normalized such that the maximal absolute value is 1. There are five categories of the interaction vertices, the first and second categories are intravalley processes, and the last three categories are intervalley processes. Intra-V Phonon stands for the phonon-mediated intravalley electron interaction, which is $10^{-3}$ times the Coulomb interaction. }
\label{fig:interactions}
\end{figure*}

The set of $\{U,J_H,V\}$ and $\{U,\tJ_H,\tV\}$ in the hot-spot model are related by Fierz identity, and the interaction strengths are related by $(\tJ_H,\tV)^\intercal = \left(\begin{smallmatrix}1/2 & 3/4 \\ -1 & 1/2  \end{smallmatrix}\right) (J_H,V)^\intercal$. As shown in \figref{fig:interactions}, one can check that the linear combination of $J_H,V$ gives rise to $\tJ_H,\tV$. Based on these considerations, we can write down the following model interaction,
\begin{equation}
    H^{\mathrm{U\tilde{V}\tilde{J}}}_{\mathrm{int}} = H_U + H_{\tV} + H_{\tJ}
\end{equation}
Compared to realistic Coulomb interaction and phonon-mediated interaction, the model interaction $H^{\mathrm{U\tilde{V}\tilde{J}}}_{\mathrm{int}}$ is short-ranged and does not take the momentum-space form factor into account. Nevertheless, as we will show in the rest of the paper, this simple interaction still captures all the essential physics of leading instabilities in the hot-spot model.

% \onecolumngrid
% \twocolumngrid

\subsection{Coulomb and phonon mediated interaction}\label{sec:coulomb}

We will also consider the more realistic Coulomb interaction and phonon-mediated interaction between electrons in RTG. The Coulomb interaction is a density-density interaction with both intra- and intervalley components,
\begin{equation}
    H_C = \frac{1}{2 A} \sum_{\tau,\tau',\boldq} V(\boldq + (\tau' - \tau) \vect{Q} ): \rho_{\tau\tau'}(\boldq) \rho_{\tau'\tau}(-\boldq):,
\end{equation}
where $::$ denotes normal ordering, $\vect{Q}$ is the momentum transfer between two valleys (upto reciprocal lattice vectors), $A$ is the sample area, $V(\boldq)=e^{2} \tanh (|\boldq| D) /(2 \epsilon |\boldq|)$ is the dual gate-screened Coulomb interaction with gate-sample distance $D$, and
\begin{equation}
    \rho_{\tau\tau'}(\boldq) = \sum_{\bk s \sigma} c_{\tau \bk s \sigma}^{\dagger} c_{\tau' \bk+\boldq s \sigma} = \sum_{\bk s \sigma} \lambda_{\boldq \sigma}^{\tau\tau'}(\bk) \psi_{\tau \bk s}^{\dagger} \psi_{\tau' \bk+\boldq s}
\end{equation}
is the Fourier component of the electron density operator, with $\lambda_{ \boldq\sigma}^{\tau, \tau'}(\bk)=\left\langle u_{\tau, \bk}\left|P_{\sigma}\right| u_{\tau', \bk+\boldq}\right\rangle$ being the sublattice projected form-factor.

In RTG, the most relevant phonon mode is the in-plane longitudinal acoustic (LA) mode. Since the primary effect of the in-plane intervalley optical mode at low momenta is to couple to electrons by distorting the AB bond in each graphene layer, the lack of any spectral weight of the electrons in the valence band on both A and B sites within a single layer implies that the coupling to such a phonon mode is strongly suppressed. (see \appref{app:phonon} for details) \footnote{A more general symmetry analysis can be found in Ref.~\cite{PhononSymmetry}}. Starting from the Fröhlich Hamiltonian and neglecting retardation effects, we can obtain the electron-electron interaction mediated by LA phonons as
\begin{equation} \label{eq:H_P}
    H_{P} = - \frac{1}{2 A} \sum_{\tau, \tau', \boldq} \frac{1}{2 \rho \omega_{\boldq}^2} : \eta(\boldq) \eta(-\boldq):
\end{equation}
where $\omega_{\boldq} = c |\boldq|$ is the phonon energy, and
\begin{equation}
    \eta(\boldq) = \sum_{\tau \bk s \sigma} M_{\bk,\boldq} \left(\lambda_{\boldq \sigma}^{\tau\tau}(\bk)\right)^* \psi_{\tau \bk+\boldq s}^{\dagger} \psi_{\tau \bk s}
\end{equation}
with $M_{\bk,\boldq} = \alpha |\boldq|$ being the electron-phonon coupling matrix element. Taking $\alpha = \SI{3.25}{\eV}$, $c_{LA} = \SI{21.2E3}{m/s}$, and $\rho = \SI{7.6E-7}{\kg/m^2}$ from \textit{ab initio} calculations \cite{TBDFT,DFT}, we find that the phonon mediated interaction is parametrically much weaker than the Coulomb interaction.

If we consider projecting the realistic Coulomb interaction and phonon-mediated interaction between electrons to the interaction basis discussed earlier, the Coulomb interaction is similar to the sum of a repulsive density-density interaction (both intravalley and intervalley) and a small ferromagnetic spin-triplet IVC Hund's coupling, i.e. $H_C \sim H_U +H_V+H_{\tJ}$, while the phonon mediated interaction is similar to the sum of an attractive intravalley and intervalley density-density interaction, i.e. $H_P \sim -H_U - H_V$ (\figref{fig:interactions}). However, the strength of the phonon-mediated interaction is relatively small and it barely modifies the density-density interaction part of the Coulomb interaction, therefore, we neglect the effect of phonon-mediated interaction in the following analysis.

\section{Renormalization group analysis}\label{sec:RG}
To explore the interaction effects, we follow the Wilsonian renormalization group approach, starting with the Hamiltonian in \eqnref{eq:hotspotham} with bare four-fermion interactions at UV cutoff $\LUV$, and gradually integrating out the high energy modes to get the effective action at the running energy scale $\Lambda$.
The bare interactions will be dressed by one-loop corrections and their strengths depend on the running energy scale $\Lambda$. At a certain critical energy scale $\Lambda_c$, some interaction may diverge and lead to corresponding instabilities, or if no interaction diverges, then the system remains in Fermi liquid phase. To implement the RG, we consider patches around the hot-spots as shown in \figref{fig:fsandnesting} (h), assuming that the main contribution of the low-energy physics comes from electronic states inside the patches since the DOS near these hot-spots diverges as $\rho(\Lambda)\sim \log(\LUV/\Lambda)$.

We assume the Fermi surface is not changed under RG, namely the self-energy correction is neglected. To study different bare interactions at the UV cutoff $\LUV$, we derive the one-loop RG equation for the coefficients of the interaction vertex basis introduced in \secref{sec:int_basis}. Then the bare interactions at the UV cutoff are represented in terms of the interaction vertex basis, the coefficients are the initial condition of the RG equation.
\subsection{Renormalization group equation}\label{sec:RGE}
\begin{figure}[htbp]
    \centering
    \includegraphics[width=0.45\textwidth]{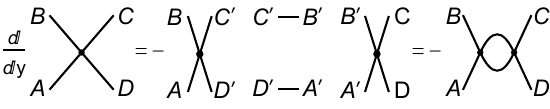}
    \caption{The interaction vertex in Majorana fermion basis is rank-4 totally antisymmetric tensor denoted by the cross with a solid black dot. The parallel two lines represent the susceptibility, which can be viewed as a tensor product of two propagators. The single one-loop diagram is equivalent to 5 one-loop diagrams in the complex fermion basis, since particle/hole are on equal footing in the Majorana basis and the projection matrices in the susceptibilities enforce the contribution from correct channels.}
    \label{fig:oneloop}
\end{figure}

It is convenient to use Majorana fermion basis $\psi_i = \chi_{i,1}+\ii \chi_{i,2}$ in the calculation ($i$ stands for the generic label of the valley, hot-spot and spin), because the interaction vertices of Majorana fermion are totally antisymmetric, it will reduce the number of Feynman diagrams we need to evaluate in the one-loop calculation. The general interaction term is expressed as $V_{ABCD}\chi_A \chi_B \chi_C \chi_D$, where the capital letter index refers to the combined index of the valley, hot-spot, spin and particle/hole indices of Majorana fermion,
\begin{equation}
    \chi_A = \begin{bmatrix}K\\K'\end{bmatrix}\otimes \begin{bmatrix}\alpha=1\sim6\end{bmatrix} \otimes \begin{bmatrix}\uparrow \\ \downarrow\end{bmatrix} \otimes \begin{bmatrix}\Re{c} \\ \Im{c} \end{bmatrix}.
\end{equation}
As shown in \figref{fig:oneloop}, the one-loop correction for the interaction vertex $V_{ABCD}$ is simply,
\begin{align}\label{eq:rge}
    &\frac{\bd V_{ABCD}(\Lambda)}{\bd \Lambda}  \nonumber\\
    &= -V_{ABC'D'}(\Lambda)\frac{\bd [\chi_\text{Maj}(\Lambda)]_{C'B';D'A'}}{\bd \Lambda}V_{A'B'CD}(\Lambda)
\end{align}
where duplicate indices are contracted automatically, and the energy scale dependent susceptibility in Majorana basis is the summation of susceptibility in particle-particle channel and particle-hole channel with projection matrices. More explicitly, the susceptibility in Majorana basis carries the indices $[\chi_\text{Maj}(\Lambda)]_{CB;DA}\equiv [\chi_\text{Maj}^{\tau\alpha,\tau'\alpha'}(\Lambda)]_{cb;da}$ where lower case letters run over spin and particle/hole indices, the susceptibility in Majorana basis is decomposed as, $\chi_\text{Maj}=\chi_\text{pp}^{\tau\alpha,\tau'\alpha'}P^\text{pp}_{cb;da}+\chi^{\tau\alpha,\tau'\alpha'}_\text{ph}P^\text{ph}_{cb;da}$, with $P^{\text{pp/ph}}_{cb;da} = [(\sigma^{00}_{cb}\otimes\sigma^{00}_{da}\pm\sigma^{02}_{cb}\otimes \sigma^{02}_{da})/2]$ and $\sigma^{ij}$ is the kronecker product of Pauli matrices $\sigma^i$ and $\sigma^j$. We note that the contraction of this single one-loop diagram is equivalent to the usual five one-loop diagrams in the complex fermion basis, because the projection matrices and totally antisymmetry of the interaction vertex in Majorana fermion basis automatically take care of the symmetry factors and contributions from various channels. This enables us to use a computer program to generate the RG equations for all the 43 linearly independent interactions more reliably.

The susceptibilities with momentum transfer relating to two hot-spots are given by,
\begin{align}
    \chi_\text{pp}^{\tau\alpha,\tau'\alpha'}(\Lambda) = \sum_{\boldp}\ &\frac{n_F(\epsilon^{\tau\alpha}_\boldp)-n_F(-\epsilon^{\tau'\alpha'}_{-\boldp})}{-\epsilon^{\tau\alpha}_\boldp-\epsilon^{\tau'\alpha'}_{-\boldp}} \nonumber\\
    &\quad \Theta(\abs{\epsilon^{\tau\alpha}_\boldp+\epsilon^{\tau'\alpha'}_{-\boldp}}-\Lambda)\\
    \chi_\text{ph}^{\tau\alpha,\tau'\alpha'}(\Lambda) = \sum_{\bk}\ &\frac{n_F(\epsilon^{\tau\alpha}_\boldp)-n_F(\epsilon^{\tau'\alpha'}_{-\boldp})}{-\epsilon^{\tau\alpha}_\boldp+\epsilon^{\tau'\alpha'}_{-\boldp}}\nonumber\\
    &\quad \Theta(\abs{\epsilon^{\tau\alpha}_\boldp-\epsilon^{\tau'\alpha'}_{-\boldp}}-\Lambda)
\end{align}
where $\Theta(x)$ is the step function, $n_F(\epsilon)$ is the Fermi-Dirac distribution, in the $0$ temperature limit, $n_F(\epsilon)\xrightarrow{T\rightarrow 0} \Theta(-\epsilon)$. The second line of each term is used to impose the infrared running cutoff $\Lambda$. More details of the derivation can be found in \appref{app:rgdetail}.

Without any nesting of the Fermi surface, the susceptibility in particle-particle channel with two VHSs at opposite momenta and different valleys always diverges as $\chi_\text{pp}^{K\alpha,K'\alpha}(\Lambda)\sim\rho(\Lambda)\log(\Lambda)$, which is the Cooper pair susceptibility. It is convenient to define the RG scale $y$ to be the Cooper pair susceptibility, such that $\frac{d\chi_\text{pp}^{K\alpha,K'\alpha}(y) }{dy}=1$ by definition. If the Fermi surface has other nesting, then the corresponding susceptibility $\chi$ satisfies $\frac{d\chi}{dy}$ is constant, otherwise, $\frac{d\chi}{dy}\sim y^{-1/2}$.

Similar to using symmetry to relate interaction vertices, the susceptibility among symmetry related pairs of VHSs should be the same. The independent susceptibilities for different displacement fields are shown in \appref{app:suscept}.

As explained at the beginning of this section, we can express the RG equation \eqnref{eq:rge} in terms of interaction vertex basis in \secref{sec:int_basis}, by $V_{ABCD}(y)=v_i(y) u^i_{ABCD}$, where $u^i_{ABCD}$ denote the fixed interaction vertex basis, and $v_i(y)$ are their corresponding coefficients (coupling strengths) which flows with the RG scale $y$.  The RG equation for the coefficients reads as,
\begin{align}
    &\frac{d v_i(y)}{d y} = -C_{i}^{jk}v_j(y) v_k(y), \label{eq:rge_coeff}\\
    &C_{i}^{jk}=u^j_{ABC'D'}\chi^\text{Maj}_{C'B';D'A'} u^{k }_{A'B'CD} g_{il}u^{l}_{DCBA}, \label{eq:ope_coeff}
\end{align}
where repeated indices are contracted, and the metric $g_{ij}$ is the \emph{inverse} of $g^{ij}=u^i_{ABCD}u^{j}_{DCBA}$. \eqnref{eq:rge_coeff} basically states the one-loop correction of the interaction vertex is given by the operator product expansion of two other interaction vertices, the OPE coefficient is given by $C_i^{jk}$ in \eqnref{eq:ope_coeff}.

The bare interaction $V_{ABCD}(y_\text{UV})\chi_A \chi_B \chi_C \chi_D$ at the UV cutoff $\LUV$ sets the initial condition for the RG equation,
\begin{equation}\label{eq:initial_cond}
    v_i(y_\text{UV})= V_{ABCD}(y_\text{UV})g_{ij}u^j_{DCBA}.
\end{equation}
Under the RG flow, if all the coupling $v_i(y)$ flow to zero, then the interaction $V$ is an irrelevant perturbation, the system remains in the Fermi liquid phase. However, if some of the couplings diverge at the critical RG scale $y_{c}$, we can stop the RG equation and obtain the infrared diverging interaction vertex $V_{ABCD}(y_c)=\sum_i v_i(y_{c})u^i_{ABCD}$. The infrared diverging interaction vertex can be viewed as a matrix and each entry corresponds to the fermion bilinear, then the eigenvector of the leading eigenvalue of such matrix is precisely the order parameter of the corresponding instability. Alternatively, the corresponding instability is measured by the fermion bilinear operator which on condensation results in the maximum energy gain for the system, more details are presented in \appref{app:insts}.

\begin{figure*}[ht]
    \centering
    \includegraphics[width=0.98
    \textwidth]{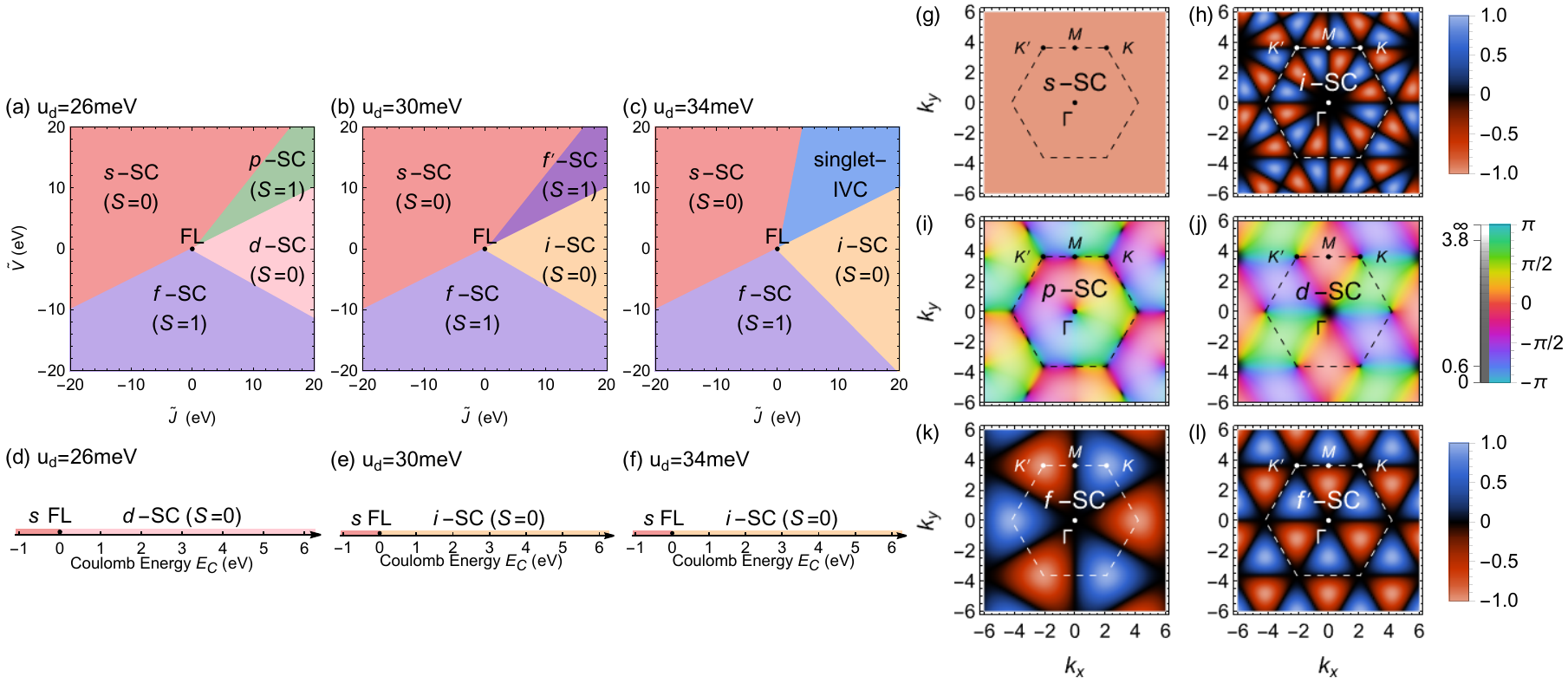}
    \caption{(a) - (c) show the phase diagrams in the $\tJ-\tV$ plane with displacement field $u_d\approx 26,30,\SI{34}{\meV}$ and $U=\tJ_H=\tJ$. The spin configurations of the superconducting order parameters are labeled by $S=0$ for spin-singlet and $S=1$ for spin-triplet. There are only superconducting phases when $u_d\approx 26,30$, but positive $\tJ_H$ and $\tV$ in (a) favor $d$-wave or $p$-wave superconductivity when $u_d\approx \SI{26}{\meV}$ due to the additional particle-particle nesting as shown in \figref{fig:fsandnesting} (d) (g). Positive $\tJ_H$ and $\tV$ favor spin singlet IVC order $\psi_{K,1}^\dagger \psi_{K',6}$ when $u_d\approx \SI{34}{\meV}$ in (c) due to the additional particle-hole nesting. (g) - (l) show various pairing symmetries appearing in the phase diagram, the pairing symmetry is with respect to the $\Gamma$ point. The red and blue dots refer to the VHSs in $K$ and $K'$ valleys respectively. $i$ and $f'$ pairing have relative $\pi$ phase rotation within the pair of VHSs. $d$-wave and $p$-wave pairing are two dimensions representations and the relative phases within the pair of VHSs are $\frac{2\pi}{3}$. (h), (l) are calculated when $u_d=\SI{30}{\meV}$, others are calculated when $u_d \approx \SI{26}{\meV}$. (d) - (f) show the phase diagrams with realistic Coulomb interaction $E_C$.}
    \label{fig:phasediag}
\end{figure*}

%\twocolumngrid

\subsection{Phase diagram}\label{sec:pd}

The bare interactions discussed in \secref{sec:interactions} will cause different instabilities under the RG flow. To better understand their behaviors, we first focus on the toy model with local interaction terms $\{H_U,H_{\tJ},H_{\tV}\}$. Since the real Coulomb interaction can be modeled as the combination of intravalley, intervalley density-density interaction, and ferromagnetic spin triplet IVC Hund's coupling, we consider the 2-dimensional phase diagram with $U=\tJ_H=\tJ$ and $\tV$ associated with the following interaction,
\begin{align}
    H_\text{int}= H_U + H_{\tJ} + H_{\tV}
    % -\tJ_H \sum_\boldq \vect{I}(\boldq)\cdot \vect{I}(-\boldq)^\dagger+U\sum_{\tau,\boldq} n_\tau(\boldq) n_\tau(-\boldq)  \nonumber\\
    % &-\tV\sum_\boldq I^0(\boldq) I^0(-\boldq)^\dagger.
\end{align}
The system with bare interaction at the UV level will flow to different stable IR phases under the renormalization group flow. We find the bare interactions in the $\tJ$-$\tV$ plane will lead to different IR phases as shown in \figref{fig:phasediag} (a)-(c) with $u_d\approx 26,30,\SI{34}{\meV}$.

As shown in \figref{fig:phasediag} (a) - (c), the different superconducting phases and singlet IVC phase compete with each other. The pairing symmetries are always defined by the winding of the pairing order parameter with respect to the $\Gamma$ point (see \figref{fig:phasediag} (g) - (l) for an illustration of  different pairing symmetries). All the phase diagrams contain spin-triplet $f$-wave superconductivity and spin-singlet $s$-wave at the bottom and top-left of the phase diagrams. These two phases are separated by a line corresponding to the change from ferromagnetic spin-Hund's coupling $J_H$ to antiferromagnetic spin-Hund's coupling $-J_H$. As discussed in \secref{sec:interactions}, the phase diagram tuning parameters $\tJ_H$ and $\tV$ are related to the intervalley spin Hund's coupling $J_H$ and the intervalley density-density interaction $V$ by a linear transformation, and $\tJ_H-\frac{3}{2}\tV>0$ corresponds to the ferromagnetic spin Hund's coupling, which covers the spin-triplet $f$-wave superconducting phase. On the other hand, $\tJ_H-\frac{3}{2}\tV<0$ favors spin-singlet $s$-wave pairing due to the antiferromagnetic spin Hund's coupling.

For the phases to the right of the phase diagrams, it is rather unusual that the positive $\tJ$ with small $\tV$, corresponding to the ferromagnetic IVC Hund's coupling, favors spin-singlet pairing $i$-wave or $d$-wave. As explained in \cite{ivcchatterjee2021rtg}, and in analogy to the cases in cuprates and magic angle twisted bilayer graphene \cite{scalapino1995case,isobe2018unconventional,you2019superconductivity}, the \emph{ferromagnetic} IVC Hund's coupling $\tJ_H$ actually promotes spin \emph{antiferromagnetic} pairing $H_{\Delta} \sim - \Delta_{-\bk}^{ \dagger} \Delta_{\bk}$ with $\Delta_{\bk} = \ii c_{\vect{K}, \bk, \uparrow} c_{\vect{K}',- \bk, \downarrow}$ and a sign changing form factor $\Delta_{\bk} = - \Delta_{-\bk}$, which in turn leads to \emph{spin-singlet} superconductivity with phase difference among these points, such as $i$-wave pairing in \figref{fig:phasediag} (b), (c) and $d$-wave pairing in \figref{fig:phasediag} (a).
If the singlet IVC coupling $\tV$ increases, it then favors spin-triplet pairing with phase difference among the hot-spots as in \figref{fig:phasediag} (a), (b) or spin singlet IVC ordering as in \figref{fig:phasediag} (c). The order parameters of these superconducting states indeed have different phase structures among the VHSs, for example, the spin-triplet $f'$-wave pairing as illustrated in \figref{fig:phasediag} (l), to be distinguished from $f$-wave pairing by the relative $\pi$ phase rotation within the pair of VHSs, say $\bK_{K1},\bK_{K4}$. The order parameters are summarized in \tabref{tab:orderp}.

The phase diagram of $u_d\approx \SI{26}{\meV}$ in \figref{fig:phasediag} (a) is different from the generic case with $u_d =\SI{30}{\meV}$ in \figref{fig:phasediag} (b) due to the additional particle-particle nesting as shown in \figref{fig:fsandnesting} (g). The additional diverging susceptibility turns the $f'$-wave and $i$-wave pairing to the spin-triplet $p$-wave and spin-singlet $d$-wave pairing as illustrated in \figref{fig:phasediag} (i) (j). This type of nesting is also akin to the case with an annular Fermi surface, because the particle-particle nesting has a momentum transfer different from the Cooper-pairing channel. From our hot-spot model, we cannot determine whether the pairing symmetry is nodal $p$-wave or the chiral $p_x+\ii p_y$, since the hot-spot model does not have the full information of the Fermi surface. We find that $p_x$ and $p_y$ are degenerate, and thus the superconducting order parameter can be any linear combination of these two. However, the chiral $p_x+\ii p_y$ would be the most energetically favorable since it is fully gapped (non-nodal), a similar situation happens for $d$-wave pairing. The transition between the spin-triplet $p$-wave and spin-singlet $d$-wave is also driven by the competition between the ferromagnetic IVC Hund's coupling $H_{\tJ}$ and singlet IVC coupling $H_{\tV}$.

When $u_d\approx \SI{34}{\meV}$, the Fermi surface has additional particle-hole nesting as in \figref{fig:fsandnesting} (i). This additional diverging susceptibility in the particle-hole channel will cause charge ordering, and indeed the IR phase has spin-singlet IVC order that physically corresponds to a charge-density wave \cite{ivcchatterjee2021rtg} instead of spin-triplet pairing as shown in the blue region of \figref{fig:phasediag} (c). The spin-singlet IVC order parameter is $\psi_{K1}^\dagger \sigma^0 \psi_{K'6}$ and other symmetry related fermion bilinears, where $\sigma^0$ is the $2\times 2$ identity matrix acting on spin indices.

\begin{figure}[hbtp]
    \centering
    \includegraphics[width=0.25\textwidth]{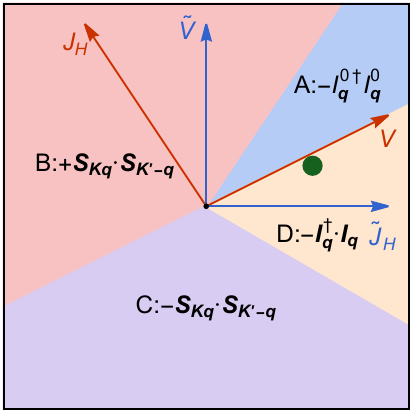}
    \caption{The different regions are dominated by the interaction terms shown in the figure, and the interaction strengths are normalized to 1. The red (purple) region is dominated by antiferromagnetic (ferromagnetic) spin Hund's coupling. The blue region is dominated by spin-singlet IVC Hund's coupling and the orange region is dominated by spin-triplet IVC Hund's coupling. The realistic Coulomb interaction (the green dot) resides below the $V$ axis in the ferromagnetic spin-triplet IVC Hund's coupling dominated region. }
    \label{fig:scheme}
\end{figure}

The competition between spin Hund's and IVC Hund's coupling almost entirely determines the phase diagram in \figref{fig:phasediag} (a)-(c). All possible pairing symmetries in RTG are shown in \figref{fig:phasediag} (g)-(l), among which the $p$-wave and $d$-wave can be either chiral or nodal \cite{GrapheneSCSym}. Which pairing symmetry shows up in the system is dominated by different Hund's terms decomposed from the bare interactions in different parameter regimes. We summarize the dominating interaction and possible resultant phases in different parameter regimes in \figref{fig:scheme} and \tabref{tab:orderp}.

In \figref{fig:scheme}, the horizontal and vertical axes in blue are ferromagnetic spin-triplet $\tJ_H$ and spin-singlet $\tV$ IVC Hund's coupling, while the red ones are intervalley density-density interaction $V$ and antiferromagnetic spin Hund's coupling $J_H$. These two sets of axes are related by the linear transformation discussed in \secref{sec:interactions}. Region B (see \figref{fig:scheme}) is dominated by the antiferromagnetic spin Hund's coupling, which favors spin-singlet pairing with a sign-preserving form factor. Then the only possible pairing symmetry once the system enters the superconducting phase is $s$-wave. A similar argument can be applied to Region C, which is dominated by ferromagnetic spin Hund's coupling and favors the spin-triplet $f$-wave pairing.

\begin{table*}[ht]
    \renewcommand*{\arraystretch}{1.6}
    \centering
    \begin{tabular}{ccccc}
        \hline \hline
       Region & Interaction & Phase & Order parameter & Form factor \\ \hline
       \multirow{2}{*}{A} & \multirow{2}{*}{$-\sum_{\boldq}I^{0\dagger}_\boldq I^0_\boldq$}  & spin-triplet SC & $f(\vect{k})c_{\bK,\bk}\ii \sigma^2 \sigma^{1,2,3} c_{\bK',-\bk}$ & $p,f'$-wave\\\cline{3-5}
       & & spin-singlet IVC & $c_{\bK',\bk+\boldq}^\dagger \sigma^0 c_{\bK,\bk}$ & $\boldq=\vect{Q}_{1-6}$\\ \hline
       B & $+\sum_{\boldq}\vect{S}_{\bK \boldq}\cdot \vect{S}_{\bK' -\boldq}$ & spin-singlet SC & $f(\vect{k})c_{\bK,\bk}\ii \sigma^2 c_{\bK',-\bk}$ & $s$-wave \\ \hline
       C & $-\sum_{\boldq}\vect{S}_{\bK \boldq}\cdot \vect{S}_{\bK' -\boldq}$ & spin-triplet SC & $f(\vect{k})c_{\bK,\bk}\ii \sigma^2 \sigma^{1,2,3} c_{\bK',-\bk}$ & $f$-wave \\ \hline
       D & $-\sum_{\boldq}\vect{I}^{\dagger}_\boldq \cdot  \vect{I}_\boldq$  & spin-singlet SC & $f(\vect{k})c_{\bK,\bk}\ii \sigma^2 c_{\bK',-\bk}$ & $d,i$-wave\\\hline \hline
    \end{tabular}
    \caption{List of regions and dominated interactions in \figref{fig:scheme}. The dominated interactions will develop different instabilities depending on the displacement field $u_d$ as shown in \figref{fig:phasediag}. The phases and their order parameters are summarized in the 3rd to 5th column. The operators of each order parameter should be combined with their hermitian conjugate and summed over $\bk$. Pauli matrices are acting on the spin indices of the fermion operators. $\vect{Q}_{1-6}$ denotes the momentum that connects hot spots 1 and 6 as well as other symmetry-related momenta.}
    \label{tab:orderp}
\end{table*}

On the right side of the diagram, the phases are separated by the positive $V$ axis ($J_H=0$). Region A is dominated by ferromagnetic spin-singlet IVC Hund's coupling, which interestingly favors spin-triplet pairing superconductivity with sign-changing form factors (or spin-singlet IVC ordering). In this case, $p$-wave and $f'$-wave pairing are both possible from a symmetry perspective and their relative susceptibilities determine the dominant instability. We identified both of them in different nesting structures as shown in \figref{fig:phasediag} (a)-(b). Finally, the most experimentally relevant Region D is dominated by the ferromagnetic spin-triplet IVC Hund's coupling and favors spin-singlet pairing superconductivity with sign-changing form factors (or spin-triplet IVC ordering). $d$-wave and $i$-wave pairing are both consistent with the symmetry and we identified both of them in \figref{fig:phasediag} (a)-(c).
Those tendencies to different orderings are further enhanced by fermion susceptibilities under the renormalization group flow and therefore yield the IR phase diagrams in \figref{fig:phasediag}.

\figref{fig:phasediag} (d)-(f) shows the phase diagram with realistic Coulomb interaction. The realistic Coulomb interaction is roughly the combination of intravalley and intervalley density-density interactions, and a relatively small ferromagnetic spin-triplet IVC Hund's coupling. The repulsive Coulomb interaction schematically resides at the green point in \figref{fig:scheme} which is below the $V$ axis and in the ferromagnetic spin-triplet IVC Hund's coupling dominated region (the yellow region). We note that the small portion of ferromagnetic spin-triplet IVC Hund's coupling in the Coulomb interaction is crucial for the system to reside in the yellow region. Then the repulsive Coulomb interaction would favor spin-singlet $d$-wave pairing for $u_d\approx \SI{26}{\meV}$ and spin-singlet $i$-wave pairing for $u_d >\SI{30}{\meV}$. If the density-density interaction is rendered attractive, then it favors spin-singlet $s$-wave pairing for all three cases. More details are summarized in \tabref{tab:orderp}.

\section{Conclusion}\label{sec:concl}
In this paper, we have used hot-spot RG to study the correlated electronic behavior of hole-doped RTG in the presence of VHSs on the Fermi surface. The DOS diverges near the VHSs and the large DOS will renormalize the interactions and cause various instabilities to symmetry-broken phases at low temperatures. The resulting phases depend on the initial interactions and the nesting structures of the Fermi surfaces. In the experiment parameter regime, we study different nesting structures of the Fermi surface with 6 VHSs (hot-spots) in each valley. As discussed in \secref{sec:nesting}, generic Fermi surfaces (e.g., when the displacement field $u_d=\SI{30}{\meV}$)
%generic Fermi surfaces only have diverging Cooper pairing channel as the case when the displacement field $u_d=\SI{30}{\meV}$, then the electrons tend to form various superconducting orders.
feature divergence only in the Cooper pairing channel, and the resultant instability is towards various superconducting phases.
When the displacement field is tuned to $u_d\approx \SI{26}{\meV}$, the Fermi surface has additional particle-particle nesting which will enhance the superconducting order and affect the dominant pairing channel. Additionally, when tuning the displacement field $u_d\approx \SI{34}{\meV}$, the 6 VHSs in each valley have additional $C_6$ symmetry, and there is an additional particle-hole nesting which favors charge ordering.

We investigated the low temperature electronic phases under different types of local interactions, especially the spin-Hund's coupling and the spin-singlet or spin-triplet IVC Hund's coupling. We find the ferromagnetic (antiferromagnetic) spin-Hund's coupling favors spin-triplet $f$-wave (spin-singlet $s$-wave) pairing. The ferromagnetic spin-triplet IVC Hund's coupling favors spin-singlet $d$-wave pairing for $u_d\approx \SI{26}{\meV}$ and $i$-wave pairing for the other two nesting structures. As discussed in \secref{sec:pd}, the ferromagnetic spin-triplet IVC Hund's coupling promotes antiferromagnetic fluctuations among the antipodal points on the Fermi surface, and leads to the spin-singlet superconductivity with phase difference among the hot-spots \cite{ivcchatterjee2021rtg,scalapino1995case,isobe2018unconventional,you2019superconductivity}. The ferromagnetic spin-singlet IVC Hund's coupling favors either spin-singlet IVC ordering if the interaction is enhanced by the particle-hole nesting as when $u_d\approx \SI{34}{\meV}$, or spin-triplet $p$-wave pairing for $u_d\approx \SI{26}{\meV}$ and $f'$-wave pairing for $u_d=\SI{30}{\meV}$.

Armed with a thorough understanding of ordering instabilities caused by local interactions, we considered realistic Coulomb interaction and phonon-mediated electronic interaction in RTG. The realistic Coulomb interaction is like the combination of repulsive density-density interaction (both intravalley and intervalley) and ferromagnetic spin-triplet IVC Hund's coupling, and it favors spin-singlet superconductivity with different phases of the order parameters among the VHSs, such as $d$-wave for $u_d\approx \SI{26}{\meV}$ and $i$-wave for the other two nesting structures with $u_d\ge \SI{30}{\meV}$. The phonon-mediated electron interaction in RTG is driven mainly by acoustic phonons and is roughly the sum of attractive intra- and intervalley density-density interaction, but its strength is significantly smaller than the Coulomb interaction (by an approximate factor of $10^{-3}$). Therefore, it barely modifies the density-density interaction vertex that is almost solely determined by the Coulomb interaction, and we can safely neglect its minuscule effect on the phase diagram.
%\textcolor{red}{Maybe also state that only acoustic phonons matter, not sure community appreciates that yet?}

The results of our analysis are in good agreement with the experimental phenomenology of the dominant superconducting phase (SC1) of RTG, which appears to be a spin-singlet.
Our calculations further identify two distinct spin-singlet superconducting phases ($d$-wave and $i$-wave) as relevant candidates for SC1.
The $d$-wave superconducting phase is expected to be fully gapped and chiral, and should therefore exhibit spontaneous orbital magnetization and edge currents that can be detected by scanning nano-SQUIDs \cite{Furusaki2001}.
The $i$-wave superconductor is nodal with gapless quasiparticle excitations and might be diagnosed using low-temperature thermal transport \cite{DurstLee}.
In addition, single spin-qubit based current noise spectroscopy \cite{CD2021,DC2021} may also be used to distinguish these different superconducting phases in experiments.

It is worthwhile to critically analyze our approach to correlated physics in RTG, by comparing with related few-layered graphene-based heterostructures such as twisted bilayer graphene (TBG) \cite{cao2018moire1,cao2018moire2}.
While both materials show strong electronic correlations due to enhanced density of states, such as iso-spin polarized phases and superconductivity, there are significant differences that should be kept in mind.
First, unlike moir\'e graphene which features interaction induced insulators at certain integer fillings of moir\'e mini-bands, RTG always has Fermi pockets away from the single-particle gap at charge neutrality.
Second, resistivity measurements point to the presence of a \textit{strange metal} phase above superconducting $T_c$ in TBG \cite{Polshyn2019,PhysRevLett.124.076801,PhysRevLett.124.186801}, whereas quantum oscillations in RTG indicate that the parent state of the superconductor is a simple Fermi liquid \cite{trg_exp2021half,trg_exp2021superconductivity,AndreaBilayer}.
Taken together, these considerations imply that superconductivity is likely a weak-coupling instability in RTG, and justify our focus on low-energy electrons at the Fermi surface.
In addition, superconductivity in RTG requires a strong displacement field, that results in sublattice polarization of the electronic wave-functions in the low-energy bands.
As we showed earlier, this causes the intervalley scattering by phonon modes to be highly suppressed in RTG, in contrast to TBG where these modes mediate an attractive interaction between electrons and may play an important role in pairing \cite{PhysRevLett.122.257002,doi:10.1073/pnas.2107874118,PhysRevLett.121.257001}.
This observation bolsters the case for an all-electronic mechanism of superconductivity in RTG, as we studied in this work.

% Finally, we want to comment on a few differences between RTG in this work and the widely studied twisted bilayer graphene (TBG). The most obvious one is the presence of band gap in TBG due to moiré superlattice \cite{cao2018moire1,cao2018moire2}, while RTG always has a Fermi surface away from bulk gap, therefore we focus on the effect of Fermi surface structure in this work. The existence of insulating state adjacent the superconducting phase in TBG also raises the question whether the doped insulator above the superconducting transition temperature $T_c$ is a simple Fermi liquid \cite{Polshyn2019,PhysRevLett.124.076801,PhysRevLett.124.186801}, whereas in RTG there is no insulating state away from charge neutrality \cite{trg_exp2021half,trg_exp2021superconductivity,AndreaBilayer}. In addition, we show in this work that the electron wavefunctions in RTG are sublattice polarized, and therefore the intervalley phonon couplings have to vanish, whereas in TBG the phonon couplings could play a more important role in intervalley superconductivity \cite{PhysRevLett.122.257002,doi:10.1073/pnas.2107874118}. Last but not least, whether the TBG superconductor is spin-singlet or triplet is still under debate \cite{2204.12579}, while for ABC there is good experimental evidence that SC1 is a spin-singlet, while SC2 is spin-polarized \cite{trg_exp2021half,trg_exp2021superconductivity,AndreaBilayer}.

Beyond the relevance of our computation to the physics of RTG, from a purely theoretical standpoint we have simultaneously devised a systematic way to represent the interaction vertices and perform the one-loop renormalization group analysis for the interactions by using the interaction vertex basis. This method essentially transforms a highly technical RG calculation to relatively simple tensor contractions which can be efficiently implemented on a computer. The use of the Majorana fermion basis enables us to represent the four fermion interactions by rank-4 totally antisymmetric tensors. The one-loop renormalization group equations are obtained by evaluating 1 diagram in the Majorana basis, instead of 5 diagrams in the complex fermion basis. This automated procedure eliminates the chance of over counting symmetry factors. It is straightforward to apply to other systems that feature Van Hove singularities or can be described by hot-spot models \cite{isobe2018unconventional,lin2019chiraltbg,lin2020parquettbg,PhysRevResearch.1.033206}.

{\it Notes Added:} While completing this manuscript, related work appeared \cite{qin2022frg_trg}, which studied superconductivity in RTG using the functional renormalization group method. They identified spin-singlet $d$-wave superconductivity under Coulomb interaction at the doping level that features annular Fermi surfaces. Our work focus on a slightly different doping level that features VHSs. We find spin-singlet superconductivity with possible $d/i$-wave depending on the displacement field.
In particular, $d$-wave superconductivity occurs when $u_d \approx \SI{26}{\meV}$ which consists of additional nesting in a particle-particle channel similar to that in the annular Fermi surface.
% RTG at a different doping level that features annular Fermi surfaces using functional renormalization group method. They identified spin-singlet $d$-wave superconductivity in this system with realistic Coulomb interaction. Our work shows the spin-singlet superconductivity is favorable with possible $d/i$-wave depending on the displacement field. In particular, $d$-wave superconductivity occurs when $u_d\approx \SI{26}{\meV}$ which consists of additional nesting in particle-particle channel and this nesting structure is similar to that in the annular Fermi surface. \textcolor{red}{Any connections to this work, (added, please check)}

\begin{acknowledgments}
We acknowledge helpful discussions with Seth Whitsitt and Michael P. Zaletel. D.C.L. and Y.Z.Y. are supported by a start-up funding at UCSD.
T.W. is supported by the U.S. DOE, Office of Science, Office of Basic Energy Sciences, Materials Sciences and Engineering Division, under Contract No. DE-AC02-05-CH11231, within the van der Waals Heterostructures Program (KCWF16).
S.C. is supported by the ARO through the MURI program (grant number W911NF-17-1-0323) via M. P. Zaletel, and the U.S. DOE, Office of Science, Office of Advanced Scientific Computing Research, under the Accelerated Research in Quantum Computing (ARQC) program via N.Y. Yao.
\end{acknowledgments}

\bibliographystyle{ucsd}
\bibliography{collection}

\clearpage
\appendix
\section{Full Hamiltonian of the two-band model}\label{app:fullham}
The RTG consists of three layers of honeycomb lattices in ABC-stacking. Its electronic states near charge neutrality mainly reside on $A_1$ and $B_3$ sites, whose low-energy band structure can be described by an effective two-band model\cite{zhang2010bandtrg1,jung2013bandtrg2,ho2016bandtrg3}
\eq{\label{eq:H0}H_0=\sum_{\tau,\vect{k},s}c_{\tau\vect{k}s}^{\dagger}h_{\tau}(\vect{k})c_{\tau\vect{k}s},}
where $c_{\tau\vect{k}s}=(c_{\tau\vect{k}s A_1},c_{\tau \vect{k}s B_3})^\intercal$ denotes the electron annihilation operator. $\tau=K,K'$ labels the valley, $s=\uparrow/\downarrow$ labels the spin, and $\vect{k}$ labels the momentum deviation from the corresponding valley center. The band Hamiltonian takes the form of
$h_\tau(\vect{k})=(\epsilon_\vect{k}^\text{s}-\mu)\sigma^0+(\tau\alpha_\vect{k}^\text{ch}+\epsilon_\vect{k}^\text{tr})\sigma^1+\beta_\vect{k}^\text{ch}\sigma^2+\epsilon_\vect{k}^\text{gap}\sigma^3$,
with $\alpha_\vect{k}^\text{ch}+\ii \beta_\vect{k}^\text{ch}=\frac{v_0^3}{\gamma_1^2}(k_x+\ii k_y)^3$, $\epsilon_\vect{k}^\text{s}=(\delta+\frac{u_a}{3})-(\frac{2v_0v_4}{\gamma_1}+u_a\frac{v_0^2}{\gamma_1^2})\vect{k}^2$, $\epsilon_\vect{k}^\text{tr}=\frac{\gamma_2}{2}-\frac{2v_0v_3}{\gamma_1}\vect{k}^2$, and $\epsilon_\vect{k}^\text{gap}=u_d(1-\frac{v_0^2}{\gamma_1^2}\vect{k}^2)$. $\tau=+/-$ for $K/K'$ valley. We adopt the parameters proposed in \cite{trg_exp2021half}, namely $\gamma_0=3.1\text{eV}$, $\gamma_1=380\text{meV}$, $\gamma_2=-15\text{meV}$, $\gamma_3=-290\text{meV}$, $\gamma_4=-141\text{meV}$, $\delta=-10.5\text{meV}$, $u_a=-6.9\text{meV}$, and $v_i=\sqrt{3}a\gamma_i/2$ (for $i=0,3,4$ with $a=0.246\text{nm}$ being the lattice constant). Their physical meanings are well documented in \cite{zhang2010bandtrg1}. In particular, the parameters $u_d$ and $\mu$ are experimentally tunable by a dual-gate device\cite{trg_exp2021half,trg_exp2021superconductivity}, where $u_d=(u_1-u_3)/2$ is the potential difference between the outer layers (which is approximately proportional to the applied displacement field) and $\mu=-(u_1+u_2+u_3)/3$ is the (overall) chemical potential (assuming $u_l$ to be the electronic potential in the $l$th layer).

\section{Details of interaction vertex basis}\label{app:intervert}
The momentum conserved interaction vertices can be greatly reduced by utilizing the symmetries of the system. In our case, we consider the lattice rotation and reflection symmetry as well as the time-reversal symmetry. Since we are focusing on the Van Hove singularities in the Brillouin zone (BZ), those symmetry actions can be represented as permutations of the VHSs. If we label the VHSs in the $K$-valley by numbers from 1 to 6 and VHSs in the $K'$-valley by 7 to 12, then the symmetries are generated by,
\begin{align}
    &C_3: (1,3,2)(4,6,5)(7,9,8)(10,12,11)\\
    &M_x: (1,4)(2,6)(3,5)(7,10)(8,12)(9,11)\\
    &T: (1,7)(2,8)(3,9)(4,10)(5,11)(6,12)
\end{align}
where $M_x$ is the reflection with respect to the $k_x$-axis, $M_x: (k_x,k_y)\rightarrow (k_x,-k_y)$, $T$ is the time-reversal symmetry $T:(k_x,k_y)\rightarrow (-k_x,-k_y)$.

The momentum conserved interaction vertices can be divided into several sets where the elements are related by symmetry actions. Those sets of interaction vertices can be further grouped into 3 categories, with different valley structures,
\begin{align}
    &\psi_{(\eta^v_1\tau)\alpha_4}^\dagger \psi_{(\eta^v_2\tau)\alpha_3}^\dagger \psi_{\tau\alpha_2}\psi_{(\eta^v_1\eta^v_2\tau)\alpha_1} = \begin{gathered}\includegraphics[width=0.13\textwidth]{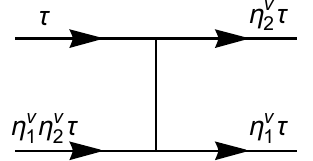}\end{gathered}
\end{align}
where $\eta^v=(1,1),(1,-1),(-1,1)$. We can further add spin indices back to the interaction vertices with $\gU(1)_s$ symmetry, then there are 9 categories of interaction vertices as shown in \figref{eq:intverts}.

In the calculation, we turn to the Majorana fermion basis, in the momentum space,
\begin{align}\label{eq:majbasis}
    &\psi_{\tau,\alpha,\boldp,s} = \chi_{\tau,\alpha,\boldp,s,1}+\ii  \chi_{\tau,\alpha,\boldp,s,2}\\
    &\psi_{\tau,\alpha,-\boldp,s}^\dagger = \chi_{\tau,\alpha,\boldp,s,1}-\ii  \chi_{\tau,\alpha,\boldp,s,2}.
\end{align}
and the Majorana fermions satisfy,
\begin{equation}
    \{\chi_i,\chi_j\} = 2\delta_{i,j}.
\end{equation}
In the Majorana fermion basis, the four fermion interaction vertices are totally antisymmetric due to the anticommutation relation of the Majorana fermions. And $\chi_A \chi_B \chi_C \chi_D$ is denoted by the cross with black dot,
\begin{align}
    &\chi_A \chi_B \chi_C \chi_D  =\begin{gathered}
        \includegraphics[width=0.08\textwidth]{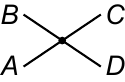}
    \end{gathered}
\end{align}
The general interaction vertices is given by rank-4 totally antisymmetric tensor with the Majorana fermion operators $V_{ABCD}\chi_A \chi_B \chi_C \chi_D$. The general rank-4 totally antisymmetric tensor can be represented by the linear combination of rank-4 totally antisymmetric tensor basis or the interaction vertex basis $u^i_{ABCD}$, where $i$ is the $i^{th}$ interaction vertex basis.

The metric of the interaction vertex basis is defined by,
\begin{equation}
    g^{ij}=\Tr{u^i_{ABCD} u^{j}_{DCBA}}
\end{equation}
where repeated indices mean contraction. Then the general rank-4 tensor of the four-fermion interaction can be expressed in terms of the interaction basis by,
\begin{equation}
    V_{ABCD}=v_i u^i_{ABCD},\quad v_i = \Tr{V_{ABCD}g_{ij}u^{j}_{DCBA}}
\end{equation}
where $g_{ij}$ is the inverse of the metric.

Note that the interaction vertex basis may not be linearly independent, one can find the linearly independent basis by Gaussian eliminating the metric $g^{ij}$ and obtaining the projection matrix $\mathsf{Proj}$ with fewer rows. Then the linearly independent basis is given by,
\begin{equation}
    \tilde{u}^i_{ABCD} = \mathsf{Proj}^{ij} u^j_{ABCD}.
\end{equation}
And the inverse of the metric is $\tilde{g}_{ij} = (\mathsf{Proj} g^{ij} \mathsf{Proj}^\intercal)^{-1}$. In the following, we will always use a linearly independent interaction vertex basis.

In the preceding analysis, we essentially view the interaction vertex as a vector, and $u^i_{ABCD}$ is the basis vector for representing the general interaction vertex. These interaction vertex basis $u^i_{ABCD}$ span the space of momentum conserved interaction vertices with respect to the discrete symmetries and at least spin $\gU(1)$ and valley $\gU(1)$ symmetries. It is easy to check that the addition of two interaction vertex basis and multiplication of scalar won't make the ``vector'' outside the space. Certain combinations of the interaction vertex basis may have larger symmetry, say, spin $\gSU(2)$ symmetry, then they span a subspace. We also note that interaction vertex basis are also closed under the contraction in the form of $\tr (u^i_{ABC'D'}u^{j}_{D'C'CD})$.

\subsection{More on symmetry actions}
In this subsection, we will discuss the interplay between symmetry and momentum conservation in the interaction vertices. As discussed previously, the VHSs are related by $C_3$ rotation symmetry, mirror reflection symmetry with respect to $k_x$-axis and time-reversal symmetry. Because of the translation symmetry, we can arrange the VHSs around the $\Gamma$ point in the first Brillouin zone. The distance of the VHSs to the $\Gamma$ point are all the same, then the symmetry actions form the discrete subgroup $\CG$ of $\gO(2)$ symmetry. If we denote a VHS as $\Ki{}$, then the other VHSs are obtained by $g_i \Ki{}$, where $g_i$ is the element in the discrete group $\CG$. The momentum conservation is given by,
\begin{align}
    g_1 \Ki{} + g_2 \Ki{}- g_3 \Ki{}-g_4 \Ki{}=0,\quad \forall \Ki{}.
\end{align}
In other word, $g_1 g_2 g_3^{-1} g_4^{-1}=\dsi $. We conclude here that momentum conservation is equivalent to the above identity of symmetry actions.

The states of the Hamiltonian at the VHSs are transformed under these symmetry actions. For our two-band model, the states are in general $\ket{\tau\alpha}=(z,a)^\intercal$ under gauge fixing, where $a$ is a fixed real number for all the states with the same energy and $z$ is a complex number depending on the valley and VHS. Then the symmetry actions on the states are,
\begin{align}
    &C_3: \ket{\tau\alpha}\rightarrow \ket{\tau\alpha'}= \ket{\tau\alpha},\\
    &M_x: \ket{\tau\alpha}\rightarrow \ket{\tau\alpha'}=\ket{\tau\alpha}^*,\\
    &T: \ket{\tau\alpha}\rightarrow \ket{\tau'\alpha}=\ket{\tau\alpha}^*
\end{align}
The symmetry actions effectively only act on the first component of the states,
\begin{align}
    &C_3:z\rightarrow z,\  M_x: z\rightarrow z^*,\  T: z\rightarrow z^*
\end{align}
One can further check that the form factors $\lambda^{\tau\alpha,\tau'\alpha'}_a=\bra{\tau\alpha}P_a \ket{\tau'\alpha'}$ in the four-fermion interactions are real as the consequence of momentum conservation,
\begin{align}
    &\lambda^{\tau_4\alpha_4,\tau_1\alpha_1}_1 \lambda^{\tau_3\alpha_3,\tau_2\alpha_2}_1 = (g_4 z)^*(g_1 z)(g_3 z)^*(g_2 z)\nonumber\\
    =&g_4^{-1}z^* g_1 z g_3^{-1} z^* g_2 z \label{eq:formfactor}
\end{align}
for elements that add phase to $z$, $g_i$ can be pulled out and by the identity $g_1 g_2 g_3^{-1} g_4^{-1}=\dsi$ from momentum conservation, $g_4^{-1}g_1 g_3^{-1} g_2 z^* z  z^*  z=z^* z  z^*  z \in \IR$. For elements correspond to mirror or time reversal symmetry that make $z$ complex conjugate, the identity $g_1 g_2 g_3^{-1} g_4^{-1}=\dsi $ implies $g_i$ with $i=(1,4),(2,3),(1,3),(2,4),(1,2,3,4)$ can be nontrivial elements, however, they come in pairs and act on $z,z^*$, therefore, the expression has the same reality as $z^*zz^*z$.

\section{Details of the Phonon Mediated Interaction}

To obtain the phonon-mediated interaction, we start with the Fröhlich Hamiltonian,
\begin{equation}
\begin{aligned}
    H&=\sum_{\tau \bk s} \psi_{\tau\bk s}^\dagger \epsilon^{\tau}_{\bk} \psi_{\tau \bk s} + \sum_{\lambda \boldq} \hbar \omega_{\lambda \tilde{\boldq}}\left(b_{\lambda \tilde{\boldq}}^{\dagger} b_{\lambda \tilde{\boldq}} +\frac{1}{2}\right) + H_{ep}\\
    H_{ep} &= \sum_{\lambda \tau \tau' \bk \boldq \sigma \sigma' s} c_{\tau' \bk+\boldq s \sigma'}^{\dagger} g_{\bk, \boldq}^{\lambda \tau \tau' \sigma \sigma'} c_{\tau \bk s \sigma}\left(b_{\lambda\tilde{\boldq}}+b_{\lambda,-\tilde{\boldq}}^{\dagger}\right)
\end{aligned}
\end{equation}
where $\tilde{\boldq} = \boldq+(\tau'-\tau)\bK$ is the physical momentum transfer, $\hbar \omega_{\lambda\boldq}$ is the phonon energy, and
$g_{\bk, \boldq}^{\lambda \tau \tau' \sigma \sigma'}$ is the electron-phonon coupling coefficient. Unlike in monolayer graphene, the valance band of RTG has spectral weight concentrated on the $B_3$ site, thus the electron-phonon coupling term can be approximated by
\begin{equation}
    H_{ep} = \sum_{\lambda \tau \tau' \bk \boldq s} \left(\lambda_{\boldq B_3}^{\tau\tau'}(\bk)\right)^* \tilde{g}_{\bk, \boldq}^{\lambda \tau \tau'} \psi_{\tau' \bk+\boldq s}^{\dagger}  \psi_{\tau \bk s}\left(b_{\lambda\tilde{\boldq}}+b_{\lambda,-\tilde{\boldq}}^{\dagger}\right)
\end{equation}
where $\tilde{g}_{\bk, \boldq}^{\lambda \tau \tau' } = g_{\bk, \boldq}^{\lambda \tau \tau' B_3, B_3}$, which suggests that only intra-sublattice phonon scattering is relevant to the low energy physics of RTG. After integrating out phonons, we get the familiar phonon-mediated interaction,
\begin{equation}
\begin{aligned}
    H_{P}&= \sum_{\lambda \tau \tau' \bk \boldq s} \left | \lambda_{\boldq B_3}^{\tau\tau'}(\bk) \right |^2 \tilde{g}_{\bk, \boldq}^{\lambda \tau \tau' } \tilde{g}_{\bk', -\boldq}^{\lambda \tau' \tau } \frac{\hbar \omega_{\lambda \tilde{\boldq}}}{\big(\varepsilon^{\tau'}_{\bk^{\prime}}-\varepsilon^{\tau}_{\bk^{\prime}-\boldq}\big)^{2}-\hbar^2\omega_{\lambda \tilde{\boldq}}^{2}}\\
    & \qquad \qquad \qquad \times : \psi_{\tau' \bk+\boldq s}^{\dagger} \psi_{\tau\bk s} \psi_{\tau\bk^{\prime}-\boldq s'}^{\dagger} \psi_{\tau'\bk^{\prime} s'} :
\end{aligned}
\end{equation}
Restricting $\bk$ and $\bk + \boldq$ to VHSs, we always have $\varepsilon^{\tau'}_{\bk^{\prime}} = \varepsilon^{\tau}_{\bk^{\prime}-\boldq}$, then the interaction simplifies to
\begin{equation} \label{eq:H_P_raw}
\begin{aligned}
    H_{P}&= - \sum_{\lambda \tau \tau' \bk \boldq s} \frac{1}{\hbar \omega_{\lambda \tilde{\boldq}}} \left | \lambda_{\boldq B_3}^{\tau\tau'}(\bk) \right |^2 \tilde{g}_{\bk, \boldq}^{\lambda \tau \tau' } \tilde{g}_{\bk', -\boldq}^{\lambda \tau' \tau } \\
    & \qquad \qquad \qquad \times: \psi_{\tau' \bk+\boldq s}^{\dagger} \psi_{\tau\bk s} \psi_{\tau\bk^{\prime}-\boldq s'}^{\dagger} \psi_{\tau'\bk^{\prime} s'} :
\end{aligned}
\end{equation}
Within the hot-spot model, the phonon-mediated interaction is always attractive in the Bloch electron basis.

We can further write $\tilde{g}_{\bk, \boldq}^{\lambda \tau \tau' }$ in terms of the phonon scattering matrix element $M^{\lambda \tau \tau'}_{\bk,\boldq}$,
\begin{equation} \label{eq:matrix_element}
    \tilde{g}_{\bk, \boldq}^{\lambda \tau \tau' }=\sqrt{\frac{\hbar}{4 A \rho \omega_{\lambda \boldq}}} M^{\lambda \tau \tau'}_{\bk,\boldq}
\end{equation}
where $\rho$ is the density of graphene and equals to $\SI{7.6E-7}{\kg/m^2}$. Rewriting \eqnref{eq:H_P_raw} in terms of $M^{\lambda \tau \tau'}_{\bk,\boldq}$, we recover the phonon mediated interaction presented in the main text (\eqnref{eq:H_P}).

\subsection{Phonon scattering matrix element in RTG} \label{app:phonon}

The momentum structure of the scattering matrix element $M^{\lambda \tau \tau'}_{\bk,\boldq}$ to the leading order in $\bk$ and $\boldq$ can be obtained by the symmetry analysis \cite{PhononSymmetry}. In this subsection, we will present a more restricted approach using the tight-binding model \cite{rana2009A1,PhononThesis}. Since only the intra-sublattice phonon scattering is relevant in RTG, we can proceed with a simplified tight-binding model that only involves the second nearest hopping between $B_3$ site,
\begin{equation}
    H_{TB} = - t_2 \sum_{nnn} \ket{\vect{R}} \bra{\vect{R}'}
\end{equation}
where $\ket{\vect{R}} = \ket{\vect{R}, B_3}$. Then the perturbative Hamiltonian due to atomic displacements is given by
\begin{equation}
    \dv{H_{TB}}{\vect{e}_{\tilde{\boldq}}^{\lambda}} = \sum_{nnn} \sum_m \dv{t_2}{b} \frac{\vect{d}_{m} \cdot \vect{e}_{\tilde{\boldq}}^{\lambda}(\vect{R})}{\sqrt{3} b} \left(1-e^{\ii \tilde{\boldq} \cdot\left(\vect{R}^{\prime}-\vect{R}\right)}\right)\ket{\vect{R}} \bra{\vect{R}'}
\end{equation}
where $\vect{e}_{\tilde{\boldq}}^{\lambda}(\vect{R}) = \tilde{\vect{e}}_{\tilde{\boldq}}^{\lambda} e^{\ii \tilde{\boldq} \cdot \vect{R}}$ is the atomic displacement of momentum $\tilde{\boldq}$, $b$ is the graphene bond length, and $\vect{d}_{m}$ are six position vectors pointing from atom $B_3$ to six nearest $B_3$ atoms. The matrix element $M^{\lambda \tau \tau'}_{\bk,\boldq}$ can be obtained from the perturbative Hamiltonian,
\begin{equation}
\begin{aligned}
    M^{\lambda \tau \tau'}_{\bk,\boldq} &= \Braket{\tau' \bk + \boldq | \dv{H_{TB}}{\vect{e}_{\tilde{\boldq}}^{\lambda}} | \tau \bk}\\
    &= \sum_m \dv{t_2}{b} \frac{\vect{d}_{m} \cdot \tilde{\vect{e}}_{\tilde{\boldq}}^{\lambda}}{\sqrt{3} b} \left(1-e^{\ii \tilde{\boldq} \cdot \vect{d}_{m}}\right) e^{\ii (\tau \vect{K} + \bk) \cdot \vect{d}_{m}}
\end{aligned}
\end{equation}
Since both $\bk$ and $\boldq$ are very small compared to the Brillouin zone scale, we can expand $M^{\lambda \tau \tau'}_{\bk,\boldq}$ in $\bk$ and $\boldq$,
\begin{align}
    M^{\lambda \tau \tau'}_{\bk,\boldq} \approx - \ii \dv{t_2}{b} \sum_m \boldq \cdot \vect{d}_{m} \frac{\vect{d}_{m} \cdot \tilde{\vect{e}}_{\tilde{\boldq}}^{\lambda}}{\sqrt{3} b} e^{\ii \tau' \vect{K} \cdot \vect{d}_{m}}
\end{align}

The next step is to write down $\tilde{\vect{e}}_{\tilde{\boldq}}^{\lambda}$ for relevant phonon modes. The relevant phonon modes at small $\bk$ and $\boldq$ are the intravalley ($\tau' = \tau$) LA and TA acoustic phonon modes and the intervalley ($\tau' = -\tau$) $A_1'$, $E'$ and $A_2'$ optical phonon modes. Intravalley optical phonon modes are strongly suppressed by the optical gap compared to the acoustic modes.
\begin{equation}
\begin{gathered}
    \tilde{\vect{e}}_{\tilde{\boldq}}^{LA} = \frac{1}{|\boldq|} \begin{pmatrix} q_x\\ q_y \end{pmatrix}, \; \tilde{\vect{e}}_{\tilde{\boldq}}^{TA} = \frac{1}{|\boldq|} \begin{pmatrix} -q_y\\ q_x \end{pmatrix},\\ \tilde{\vect{e}}_{\tilde{\boldq}}^{A_1'} = \frac{1}{\sqrt{2}} \begin{pmatrix} -1\\ \ii \end{pmatrix}, \; \tilde{\vect{e}}_{\tilde{\boldq}}^{E'} = \pm \frac{1}{\sqrt{2}} \begin{pmatrix} 1\\ \ii \end{pmatrix}, \; \tilde{\vect{e}}_{\tilde{\boldq}}^{A_2'} = \frac{1}{\sqrt{2}} \begin{pmatrix} 1\\ -\ii \end{pmatrix}
\end{gathered}
\end{equation}
Plugging in $\vect{d}_m = C_6^m \vect{d}_0$, $\vect{d}_0 = (a,0)$, and $\vect{K} = (4\pi/3a, 0)$, we find
\begin{equation}
    |M^{\lambda \tau \tau'}_{\bk,\boldq}| = \frac{3 \sqrt{3} b}{2} \dv{t_2}{b} |\boldq|
\end{equation}
for $\lambda = LA$, $A_1'$, $E'$, and $A_2'$ and otherwise zero.

The deformation potential $\alpha$ of the $LA$ mode has been studies extensively using \textit{ab initio} methods \cite{TBDFT},
\begin{equation}
    \alpha = \frac{3 \sqrt{3} b}{2} \dv{t_2}{b} = \SI{3.25}{\eV}
\end{equation}
Within the tight binding approximation, it is the same for $\lambda = LA$, $A_1'$, $E'$, and $A_2'$. Then the phonon-mediated interaction can be written as
\begin{equation} \label{eq:H_P_raw2}
\begin{aligned}
    H_{P}&= - \frac{1}{4 A \rho} \sum_{\lambda \tau \tau' \bk \boldq s} \frac{\alpha^2 |\boldq|^2}{\omega_{\lambda \tilde{\boldq}}^2} \left | \lambda_{\boldq B_3}^{\tau\tau'}(\bk) \right |^2 \\
    & \qquad \qquad \times: \psi_{\tau' \bk+\boldq s}^{\dagger} \psi_{\tau\bk s} \psi_{\tau\bk^{\prime}-\boldq s'}^{\dagger} \psi_{\tau'\bk^{\prime} s'} :
\end{aligned}
\end{equation}

For acoustic phonon modes, we approximate the phonon energy by $\omega_{\lambda \tilde{\boldq}} \approx c_{\lambda} |\boldq|$, and for optical modes, $\omega_{\lambda \tilde{\boldq}} \approx \omega_{\lambda}$ with $c_{LA} = \SI{21.2E3}{m/s}$, $\omega_{A_1'} = \SI{160}{\meV}$, $\omega_{E'} = \SI{151}{\meV}$, and $\omega_{A_2'} = \SI{124}{\meV}$ from \textit{ab initio} methods \cite{DFT}.
% From these numbers, it is clear that all intervalley modes are strongly suppressed by the large optical phonon gap

\section{Details of the Renormalization Group Approach}\label{app:rgdetail}
\subsection{Free fermion propagator}
One complication in the renormalization group analysis is to enumerate all the diagrams in particle-particle channel and particle-hole channel with correct symmetry factors. This is due to the discrepancy in complex fermion creation and annihilation operators and the complication can be overcome by using the Majorana fermion basis. Moreover, the $n$-fermion operator can be expressed as a totally antisymmetric rank $n$ tensor in Majorana fermion basis, this property largely reduces the number of diagrams that we need to sum over. For the 1-loop correction of 4-fermion interaction terms, there are 5 diagrams in complex fermiom basis correspond to only 1 diagram when using the Majorana fermiom basis.

We define the Majorana fermion at each Van Hove singularity as in \eqnref{eq:majbasis}. Note that the Majorana fermion may not correspond to the real fermions in the system, one may view this as rewriting that facilitates us to derive the renormalization group equations. The Hamiltonian in the Majorana fermion basis is thus,
\begin{equation}
    H=\sum_{\tau,\alpha,\boldp,s}\sum_{i,i'} \chi_{\tau,\alpha,-\boldp,s,i} [\epsilon^{\tau\alpha}_\boldp\sigma^2_{i,i'}] \chi_{\tau,\alpha,\boldp,s,i'},
\end{equation}
where $\sigma^2$ is the Pauli $y$-matrix acting on the real/imaginary part of the Majorana fermion operators. The free fermion propagator is readily defined as,
%= [\ii \omega-h_{\tau,\alpha,\boldp}]^{-1}
\begin{align}\label{eq:freeprop}
    &[G_0^{\tau,\alpha}]_{s,s';i,i'}(\boldp)=-\langle \chi_\boldp \chi_{-\boldp}^\intercal\rangle  \nonumber\\
    &= \sum_{j=\pm1} \frac{P^j}{\ii \omega - j \epsilon^{\tau\alpha}_\boldp},\quad P^j = \sigma^0\otimes \frac{\sigma^0+j \sigma^2}{2}.
\end{align}
\subsection{Interaction vertices}
In the following, we group the valley and hot-spot indices into $\tau\alpha$ and group the spin and Majorana real/imaginary part indices into lower case Latin letters, like $a$, $\chi_{\tau,\alpha,\boldp,s,i}\rightarrow \chi_{\tau\alpha,\boldp,a}$. The general four-fermion interaction can be written as,
\begin{align}
    &H_\text{int} = V_{abcd}(\momind{1},\momind{2},\momind{3},\momind{4})\nonumber\\
    &\chi_{\momind{1},\ki{1},a} \chi_{\momind{2},\ki{2},b}\chi_{\momind{3},\ki{3},c}\chi_{\momind{4},\ki{4},d}.
\end{align}
The valley and hot-spot indices cannot take all values since the momentum needs to be conserved. If we denote the momentum at valley $\tau$ and hot-spot $\alpha$ as $\vect{K}_{\tau\alpha}$ and recall that $\ki{i}$ are the momentum relative to the hot-spots, then the interaction vertices requires
\begin{equation}
    %(\Ki{1}+\ki{1})+(\Ki{2}+\ki{2}) - (\Ki{3}+\ki{3})-(\Ki{4}+\ki{4}) = 0
    \Ki{1}+\Ki{2}-\Ki{3}-\Ki{4} = 0.
\end{equation}
Similar to the momentum conservation, the interaction vertices also obey spin conservation and valley conservation, for example, the process which has two spin-$\uparrow$ incoming states and two spin-$\downarrow$ outgoing states is prohibited.
% fig illustrate spin conservation, hot-spots conservation

Under these constraints, the valid interaction vertices are physical, but they are still redundant and can be related under symmetry actions on the valleys and hot-spots. If there exists a symmetry $G$ such that the hot-spots are invariant under the corresponding group transformation $g$, then the interaction vertices can be related to the representative ones,
\begin{align}
    &V_{abcd}(g(\momind{1}),...,g(\momind{4})) \nonumber\\
    &= V_{abcd}(\momind{1},...,\momind{4}) ,\forall g\in G.
\end{align}
All the representatives and their orbits under the group action form the basis for the general 4-fermion interactions, we call these interaction vertices fundamental interaction vertices. The real world 4-fermion interactions are like general vectors and can be decomposed into these fundamental interaction vertices. Therefore, the behaviour of these fundamental interaction vertices under the renormalization group can determine the behaviour of other arbitrary interaction vertices.
% way to generate these fundamental vertices

\onecolumngrid
\subsection{Renormalization group equation}\label{app:RGEdetail}
For presentation clarity, we further group the indices into capital letters, $\chi_{\tau\alpha,\boldp,a}\rightarrow \chi_A$. We consider the one loop correction to the general interaction vertices at energy cutoff $\Lambda$,
\begin{equation}
H_\text{int} = V_{ABCD}(\Lambda)\chi_A \chi_B \chi_C \chi_D.
\end{equation}
Because the different Majorana fermions are anticommuting, the interaction tensors are totally antisymmetric,
\begin{equation}
    V_{ABCD}=-V_{BACD}=-V_{ACBD}=V_{BCAD}=...
\end{equation}
The one-loop correction to the system is obtained by integrating out the high energy modes. Due to the antisymmetrization, the one-loop correction for the interaction vertex is obtained by the contraction of two vertices,
\begin{align}
    &V_{ABCD}(\Lambda) \chi_A\chi_B\chi_C\chi_D \nonumber \leftarrow  -\langle V_{ABC'D'}(\Lambda) \chi_A\chi_B\chi_{C'}\chi_{D'} V_{A'B'CD}(\Lambda) \chi_{A'}\chi_{B'}\chi_C\chi_D  \rangle  \nonumber\\
    &= -V_{ABC'D'}(\Lambda)\langle \chi_{C'}\chi_{B'}\rangle \langle \chi_{D'} \chi_{A'} \rangle V_{A'B'CD}(\Lambda)  \chi_A\chi_B\chi_C\chi_D %\\
    %&= -V_{ABC'D'}(\Lambda)[G_0(\boldp)]_{C'B'}[G_0(-\boldp+\ki{1}+\ki{2})]_{D'A'}V_{A'B'CD}(\Lambda)  \chi_A\chi_B\chi_C\chi_D
\end{align}
%wait, why not take derivative wrt V_ABCD
therefore, the increment of the interaction tensors is,
\begin{equation}\label{eq:rge1}
    \frac{\bd V_{ABCD}(\Lambda)}{\bd \Lambda} = -V_{ABC'D'}\frac{\bd [\chi_\text{Maj}(\Lambda)]_{C'B';D'A'}}{\bd \Lambda}V_{A'B'CD}
\end{equation}
where the indices appeared twice mean contraction. The bare susceptibility in Majorana basis is actually the summation of the particle-particle and particle-hole channels with projection matrices,
\begin{align}
    & [\chi_\text{Maj}(\Lambda)]_{C'B';D'A'} =[\chi_\text{Maj}^{\tau\alpha,\tau'\alpha'}(\Lambda)]_{c'b';d'a'}=[\chi_\text{pp}^{\tau\alpha,\tau'\alpha'}(\Lambda)]+[\chi_\text{ph}^{\tau\alpha,\tau'\alpha'}(\Lambda)]\\
    &[\chi_\text{pp}^{\tau\alpha,\tau'\alpha'}(\Lambda)] = \int_\boldp\ \frac{\Theta(-\epsilon^{\tau\alpha}_{\boldp})-\Theta(\epsilon^{\tau'\alpha'}_{-\boldp})}{-\epsilon^{\tau\alpha}_{\boldp}-\epsilon^{\tau'\alpha'}_{-\boldp}}\Theta(\abs{\epsilon^{\tau\alpha}_{\boldp}+\epsilon^{\tau'\alpha'}_{-\boldp}}-\Lambda)P^\text{pp}\\
    &[\chi_\text{ph}^{\tau\alpha,\tau'\alpha'}(\Lambda)] = \int_\boldp\ \frac{\Theta(-\epsilon^{\tau\alpha}_{\boldp})-\Theta(-\epsilon^{\tau'\alpha'}_{-\boldp})}{-\epsilon^{\tau\alpha}_{\boldp}+\epsilon^{\tau'\alpha'}_{-\boldp}}\Theta(\abs{\epsilon^{\tau\alpha}_{\boldp}-\epsilon^{\tau'\alpha'}_{-\boldp}}-\Lambda)P^\text{ph}.
\end{align}
where the valley and hot-spot indices in the $C,B = \tau\alpha$ and $D,A=\tau'\alpha'$ entries are the same respectively, $\Theta(x)$ is the Heaviside Theta function. Note that the integration of the momentum $\boldp$ is relative to the hot-spot and inside a small disk region.

It is relative easy to obtain the renormalization group equations for the fundamental interaction vertices from \eqnref{eq:rge1}, compared to summing over 5 one-loop diagrams when using complex fermion basis. The symmetry factor and other counting factors are automatically dealt with during the tensor contraction. Despite it is convenience in implementation, the computational cost is higher than the complex fermion basis. Since we are writing the 4-fermion interaction vertices in the Majorana basis, the dimension of each entry of the interaction tensor is doubled, therefore the interaction vertices are $2^4$ larger than those in complex fermion basis. Nevertheless, it can be fast implemented and calculated for 12 hot-spots model with spin indices on PC (Intel(R) Core(TM) i7-9700K CPU @ 3.60GHz).

\subsection{Derivation of the RG equation}
The one-loop correction of the interaction vertices $V_{abcd}(\momind{1},\momind{2},\momind{3},\momind{4})$ is,
\begin{align}
    &-\int_{\bk,\omega} V_{abc'd'}(\momind{1},\momind{2},\Pi(\bk),\Pi(-\bk+\momind{1}+\momind{2}))[G_0(\bk)]_{c'b'}\nonumber\\
    &[G_0(-\bk+\momind{1}+\momind{2})]_{d'a'} V_{a'b'cd}(\Pi(-\bk+\momind{1}+\momind{2}),\Pi(\bk),\momind{3},\momind{4})
\end{align}
where $\Pi(\bk)$ is the projection operator which projects the momentum to its nearest hot-spot. Therefore, the integration of $\bk$ is the summation of possible hot-spots with integration over the relative momentum near the hot-spots. Let $\bk=\Ki{}+\boldp$, such that $\Pi(\Ki{}+\boldp)=\tau\alpha$ and $\Pi(-\Ki{}-\boldp+\momind{1}+\momind{2})=\tau'\alpha'$, then
\begin{align}
    &-\sum_{\tau\alpha}\int_{\boldp,\omega} V_{abc'd'}(\momind{1},\momind{2},\tau\alpha,\tau'\alpha')[G_0^{\tau\alpha}(\boldp)]_{c'b'} [G_0^{\tau'\alpha'}(-\boldp)]_{d'a'} V_{a'b'cd}(\tau'\alpha',\tau\alpha,\momind{3},\momind{4})
\end{align}
where $\tau'\alpha'$ depends on $\momind{1},\momind{2},\tau\alpha$ due to the momentum conservation. The integration over continuous momentum only occurs inside the bubble, we define,
\begin{align}
    [\chi_\text{Maj}^{\tau\alpha,\tau'\alpha'}]_{c'b';d'a'}&= \int d^2\boldp d\omega\  [G_0^{\tau\alpha}(\boldp)]_{c'b'}[G_0^{\tau'\alpha'}(-\boldp)]_{d'a'} \nonumber\\
    =\int d^2\boldp d\omega\ &\sum_{j,j'}\frac{1}{\ii \omega-j \epsilon^{\tau\alpha}_{\boldp}}\frac{1}{\ii \omega-j' \epsilon^{\tau'\alpha'}_{-\boldp}}[P^j]_{c'b'}\otimes [P^{j'}]_{d'a'} \nonumber\\
    =\int d^2\boldp \ &\sum_{j,j'}\frac{n_F(-j' \epsilon^{\tau'\alpha'}_{-\boldp})-n_F(j \epsilon^{\tau\alpha}_{\boldp})}{j' \epsilon^{\tau'\alpha'}_{-\boldp}+j \epsilon^{\tau\alpha}_{\boldp}}[P^j]_{c'b'}\otimes [P^{j'}]_{d'a'} \nonumber\\
    \stackrel{T\rightarrow 0}{=} \int d^2\boldp\ &\sum_{j,j'}\frac{\Theta(j' \epsilon^{\tau'\alpha'}_{-\boldp})+\Theta(j \epsilon^{\tau\alpha}_{\boldp})-1}{j' \epsilon^{\tau'\alpha'}_{-\boldp}+j \epsilon^{\tau\alpha}_{\boldp}}[P^j]_{c'b'}\otimes [P^{j'}]_{d'a'}
\end{align}
where $n_F$ is the Fermi-Dirac distribution function, when temperature goes to 0, $n_F(x)=\Theta(-x)$, where $\Theta(x)$ is the Heaviside Theta function. The projection matrices come from the propagator in \eqnref{eq:freeprop}. In the renormalization group analysis, we introduce the infrared cutoff for the energy by,
\begin{align}
    &[\chi_\text{Maj}^{\tau\alpha,\tau'\alpha'}(\Lambda)]_{c'b';d'a'} =\int d^2\boldp\ \sum_{j,j'}\frac{\Theta(j' \epsilon^{\tau'\alpha'}_{-\boldp})+\Theta(j \epsilon^{\tau\alpha}_{\boldp})-1}{j' \epsilon^{\tau'\alpha'}_{-\boldp}+j \epsilon^{\tau\alpha}_{\boldp}}\Theta(\abs{j' \epsilon^{\tau'\alpha'}_{-\boldp}+j \epsilon^{\tau\alpha}_{\boldp}}-\Lambda)[P^j]_{c'b'}\otimes [P^{j'}]_{d'a'}. \label{eq:baresusexp}
\end{align}
Let $j'=k j$, then the summation over $j,j'$ becomes $j,k$, for $k=+1$ (or $k=-1$), the expressions besides the projection matrices stay the same. Then the $\chi_\text{Maj}$ can be further simplified to a summation of particle-particle and particle-hole bare susceptibilities with projection matrices,
\begin{align}
    &[\chi_\text{Maj}^{\tau\alpha,\tau'\alpha'}(\Lambda)]=[\chi_\text{pp}^{\tau\alpha,\tau'\alpha'}(\Lambda)]+[\chi_\text{ph}^{\tau\alpha,\tau'\alpha'}(\Lambda)]\\
    &[\chi_\text{pp}^{\tau\alpha,\tau'\alpha'}(\Lambda)] = \int_\boldp\ \frac{\Theta(-\epsilon^{\tau\alpha}_{\boldp})-\Theta(\epsilon^{\tau'\alpha'}_{-\boldp})}{-\epsilon^{\tau\alpha}_{\boldp}-\epsilon^{\tau'\alpha'}_{-\boldp}}\Theta(\abs{\epsilon^{\tau\alpha}_{\boldp}+\epsilon^{\tau'\alpha'}_{-\boldp}}-\Lambda)P^\text{pp}\\
    &[\chi_\text{ph}^{\tau\alpha,\tau'\alpha'}(\Lambda)] = \int_\boldp\ \frac{\Theta(-\epsilon^{\tau\alpha}_{\boldp})-\Theta(-\epsilon^{\tau'\alpha'}_{-\boldp})}{-\epsilon^{\tau\alpha}_{\boldp}+\epsilon^{\tau'\alpha'}_{-\boldp}}\Theta(\abs{\epsilon^{\tau\alpha}_{\boldp}-\epsilon^{\tau'\alpha'}_{-\boldp}}-\Lambda)P^\text{ph}
\end{align}
where $P^{\text{pp/ph}}_{cb;da} = [(\sigma^{0000}\pm\sigma^{0202})/2]_{cb;da}$. The renormalization group equation is,
\begin{align}
    &\frac{d V(\momind{1},\momind{2},\momind{3},\momind{4})}{d\Lambda} \nonumber\\
    =& -V_{abc'd'}(\momind{1},\momind{2},\tau\alpha,\tau'\alpha')\frac{d [\chi_\text{Maj}^{\tau\alpha,\tau'\alpha'}(\Lambda)]_{c'b';d'a'}}{d \Lambda} V_{a'b'cd}(\tau'\alpha',\tau\alpha,\momind{3},\momind{4}).
\end{align}
Note that the renormalization group equation only involves tensor contraction at each step given the bare susceptibilities calculated in \eqnref{eq:baresusexp}, the tensor contraction automatically generates the momentum conserved and symmetry allowed interaction vertices since the interaction vertex basis are momentum conserved and symmetry allowed.

\subsection{Calculation of the bare susceptibilities}\label{app:suscept}
The dispersion around the Van Hove singularities is modeled by line crossing,
\begin{equation}
    \epsilon^{\tau\alpha}_{(p \cos\theta,p \sin\theta)} = e_0 p^2(\cos{2(\theta-\phi^{\tau\alpha})\sec(\psi)-1})
\end{equation}
where $\phi^{\tau\alpha}$ is the orientation of the crossing and $\psi$ is the open angle of the crossing. The change of bare susceptibilities can be calculated in the polar coordinates via the general formula,
\begin{equation}
    \frac{d}{d\Lambda }\left(\int_{\theta _0 ( \Lambda )}^{\theta _1 (\Lambda) } f(\Lambda ,\theta ) \, d\theta \right)=f\left(\Lambda ,\theta _1 (\Lambda) \right)\frac{d\theta _1 (\Lambda) }{d\Lambda }-f\left(\Lambda ,\theta _0 (\Lambda) \right)\frac{d\theta _0 (\Lambda)}{d\Lambda }+\int_{\theta _0 (\Lambda) }^{\theta _1 (\Lambda) } \frac{df(\Lambda ,\theta )}{d\Lambda } \, d\theta.
\end{equation}
For the Fermi surface without nesting, the most diverging susceptibility is the Cooper channel, which is $\chi_{pp,0}\sim \log(\Lambda)^2$ due to the nesting and large density of states at the Van Hove singularities, other channel diverges as $\log(\Lambda)$, therefore, the system would possibly flow to superconducting phase. For Fermi surface with nesting, the other channels can also diverge as $\sim \log(\Lambda)^2$ but slower than the Cooper channel, in those cases, instabilities other than superconductivity are expected.

\begin{figure}[htbp]
    \centering
    \includegraphics[width=0.95\textwidth]{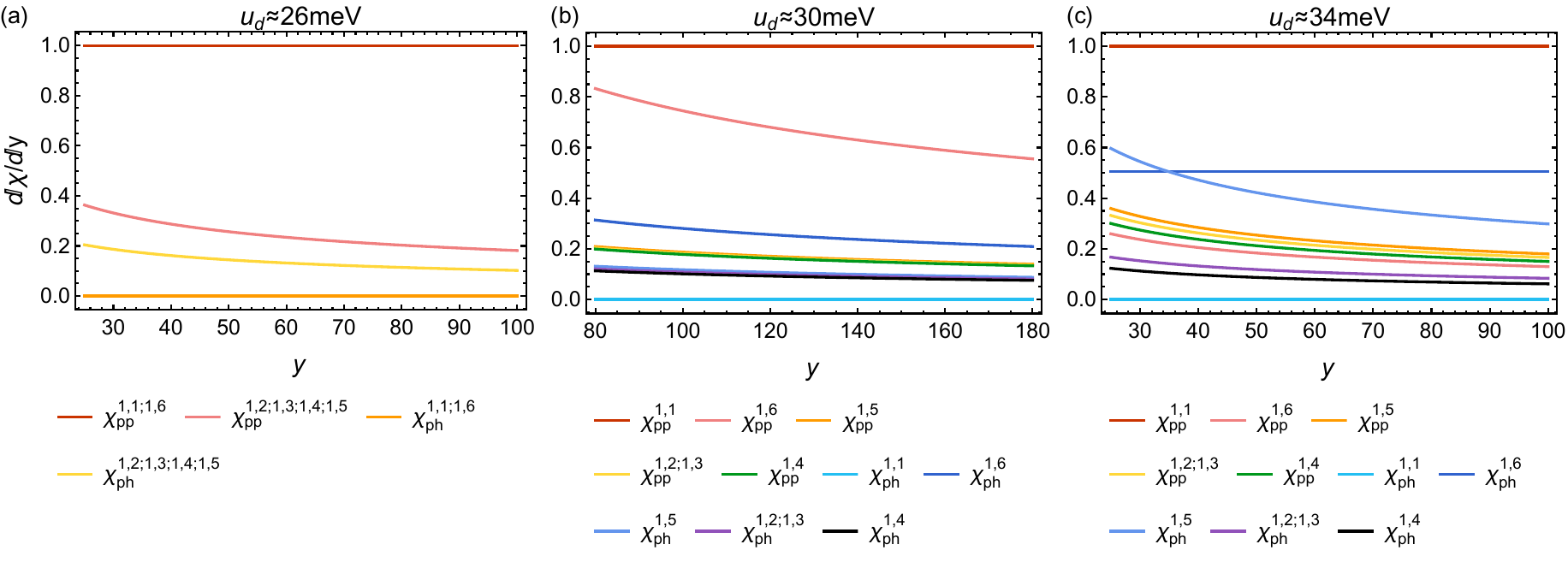}
    \caption{(a) - (c) show susceptibilities of particle-particle channels and particle-hole channels with displacement field $u_d\approx \SI{26}{\meV}$, $\SI{30}{\meV}$ and $\SI{34}{\meV}$.  }
    \label{fig:susp}
\end{figure}

Since the band structures around the VHSs are modeled by quadratic polynomials of $p_x,p_y$ which is the momentum relative to the VHSs, the Fermi surface around the VHS $\alpha$ in $K$ valley is the same as that around the VHS $\alpha$ in $K'$ valley. The susceptibilities are then labeled by $\chi^{\alpha,\alpha'}$ in \figref{fig:susp}. Due to the high symmetry, there are only 2 different processes in each channel when $u_d\approx \SI{26}{\meV}$, while there are 5 different processes in each channel when $u_d\approx 30,\SI{34}{\meV}$.

\subsection{Instabilities}\label{app:insts}
The interaction vertices will diverge or go to zero under the renormalization group flow, these correspond to relevant or irrelevant operators respectively. If no diverging interaction vertices, this means all the interactions are irrelevant, the IR phase is the Fermi liquid phase.

The instabilities of the system are characterized by the diverging interaction vertices. Once the interaction vertices diverge, we can analyze which order will form to minimize the energy. The diverged interaction vertices $\tilde{V}_{ABCD}$ can be written as,
\begin{equation}
    H_\text{int}=\tilde{V}_{ABCD}\chi_A \chi_B \chi_C \chi_D \sim -\CO_{AB}\tilde{V}_{ABCD}\CO_{CD}
\end{equation}
where $\CO_{AB} = \ii \chi_A \chi_B$. The diverged interaction vertices can be decomposed as $\tilde{V}_{ABCD} =U_{AB}^{I \dagger} \lambda^I U^I_{CD}$, where $\lambda^I$ are the eigenvalues of the matrix $\tilde{V}_{(AB)(CD)}$ where we combined the first two and last two indices. Therefore,
\begin{equation}
    H_\text{int}=-\CO_{AB}U_{AB}^{I \dagger} \lambda^I U^I_{CD}\CO_{CD}.
\end{equation}
Hence, when condensing the fermion bilinear $U^{\text{max}}_{CD} \CO_{CD}$ with the largest eigenvalue, the diverged interaction vertex will gain the most energy, and the order parameter of the leading instability is then,
\begin{equation}
   \CO = \ii U^{\text{max}}_{C,D}\chi_C \chi_D.
\end{equation}
It is often the case that several fermion bilinears have the same largest eigenvalue, they are in the subspace with remaining unbroken symmetries.

\subsection{Scaling dimension of order parameters}
The asymptotic behavior of the renormalization group equations around $y\rightarrow y_c$ is,
\begin{equation}\label{eq:raybehaviour}
    v_i = \frac{\gamma_i}{y_c-y}
\end{equation}
When plugging \eqnref{eq:raybehaviour} into the renormalization group equation \eqnref{eq:rge_coeff}, we have,
\begin{equation}
    \gamma_i = C_i^{jk} \gamma_j \gamma_k.
\end{equation}
Upon iteratively solving the equation, we find several solutions corresponding to different phases, $\gamma_i^{(\xi)}$, where $\xi$ labels the different solutions. The coefficients of the interaction vertex basis on each fixed ray behave as,
\begin{equation}
    v_i^{(\xi)} = \frac{\gamma_i^{(\xi)}}{y_c-y}
\end{equation}
The IR behavior of the system is governed by these fixed rays, since the arbitrary initial bare interactions will eventually flow to one of these fixed rays and induce corresponding instabilities.

We can further extract the scaling dimensions of the order parameters on these fixed rays. We introduce the fermion bilinear as the order parameter, $\CO =X_{AB}\chi_A \chi_B$, the vertex correction is given by,
\begin{align}
    &X_{AB}(\Lambda) \chi_A\chi_B \nonumber\\
    &\leftarrow  -\langle V_{ABC'D'}(\Lambda) \chi_A\chi_B\chi_{C'}\chi_{D'} X_{A'B'}(\Lambda) \chi_{A'}\chi_{B'}  \rangle  \nonumber\\
    &= -V_{ABC'D'}(\Lambda)\langle \chi_{C'}\chi_{B'}\rangle \langle \chi_{D'} \chi_{A'} \rangle X_{A'B'}(\Lambda)  \chi_A\chi_B %\\
    %&= -V_{ABC'D'}(\Lambda)[G_0(\bk)]_{C'B'}[G_0(-\bk+\ki{1}+\ki{2})]_{D'A'}V_{A'B'CD}(\Lambda)  \chi_A\chi_B\chi_C\chi_D
\end{align}
then the RG equation is given by,
\begin{equation}
    \frac{d X_{AB}}{dy}=-V_{ABC'D'}\frac{d [\chi_\text{Maj}]_{C'D';A'B'}}{dy}X_{A'B'}.
\end{equation}
Along these fixed rays of the RG flow, the asymptotic behavior of the corresponding fermion bilinear $\CO^{(\xi)}$ near $y\rightarrow y_c$ is,
\begin{equation}
    \frac{d\CO^{(\xi)}}{dy} = \frac{\gamma^{(\xi)}}{y_c-y}\CO^{(\xi)}
\end{equation}
whose solution is $\CO^{(\xi)}(y)\propto (y_c-y)^{-\gamma^{(\xi)}}$. In the following \tabref{tab:op_mat}, we show the scaling dimension $\gamma^{(\xi)}$ of the order parameters in different phases. The order parameter is represented by,
\begin{equation}
    \CO^{(\xi)}=\chi^\intercal M_{\alpha,\alpha'}\otimes \sigma^{\mu\nu\lambda}_{(\tau,s,a),(\tau',s',a')} \chi
\end{equation}
where $M$ is $6\times 6$ matrix acting on the indices of VHSs, $\sigma^{\mu\nu\lambda}=\sigma^\mu\otimes\sigma^\nu\otimes\sigma^\lambda$ acting on valley, spin and particle/hole indices of the Majorana fermion.

\begin{table}[hbtp]
    \centering
    \begin{tabular}{c|c|c|c|c}
        Name & Matrix form ($M\otimes \sigma^{\mu\nu\lambda}$ acts on $(\alpha,\tau,s,a)$) & Scaling dimension (at 34meV) & (30 meV) & (26meV)\\\hline
        $s$-wave SC & $diag(1,1,1,1,1,1)\otimes \sigma^{123,121}$ & $0.706$ &0.577& 0.577  \\ \hline
        $i$-wave SC & $diag(1,1,1,-1,-1,-1)\otimes \sigma^{123,121}$& $0.706$ &0.577& -\\\hline
        $p$-wave SC & $diag(-1,1,0,1,0,-1) \otimes \sigma^{233,231,203,201,213,211}$&- &-& $0.289$\\\hline
        $d$-wave SC & $diag(1,0,-1,0,-1,1) \otimes \sigma^{123,121}$& - &-& $0.408$\\\hline
        $f$-wave SC & $diag(1,1,1,1,1,1)\otimes \sigma^{233,231,203,201,213,211}$&0.497 &0.408& 0.408 \\\hline
        $f'$-wave SC & $diag(1,1,1,-1,-1,-1)\otimes \sigma^{233,231,203,201,213,211}$& - &0.408&-\\\hline
        $\text{IVC}_{1-6}^s$ & $[M|M_{1,6}=M_{2,4}=M_{3,5}=1,sym]\otimes \sigma^{200,102} $&0.349 &-&- \\ \hline
        $\text{IVC}_{1-6}^t$ & $[M|M_{1,6}=M_{2,4}=M_{3,5}=1,sym]\otimes \sigma^{210,222,230,112,120,132} $ & - &-&-
    \end{tabular}
    \caption{The first two columns show the order parameters and their matrix forms. The last three columns show the scaling dimensions of the order parameters in each phase. For different displacement fields, the phase may not exist, then there is not corresponding scaling dimension of the order parameter. }
    \label{tab:op_mat}
\end{table}

\subsection{Projected renormalization group equation}
It is hard to analyze the whole set of RG equations for interaction vertex basis (40+ first order nonlinear differential equations), we can derive the RG equations for the instabilities that we are interested in. To reproduce the phase diagram in the previous section, we only need to keep interaction vertices that correspond to the present instabilities in the \tabref{tab:op_mat}. Each order parameter corresponds to a 4-fermion interaction that can be expressed as the linear combination of interaction vertex basis,
\begin{equation}
    (\CO^i)^\dagger \CO^i = M^{i}_j u^j_{ABCD}\chi_A\chi_B \chi_C\chi_D
\end{equation}
where the superscript $i$ labels the different order parameters. The general interaction vertex in the renormalization group equation \eqnref{eq:rge1} is then $V_{ABCD}(y)=\tilde{v}_i(y) M^i_j u^j_{ABCD}$, $v_i(y)$ in \eqnref{eq:rge_coeff} is related to $\tilde{v}_i(y)$ by $v_i=(M^\intercal \tilde{v})_i$,
\begin{equation}
    \frac{d (M^\intercal \tilde{v}(y))_i}{d y} = -C_{i}^{jk}(M^\intercal \tilde{v}(y))_j(M^\intercal \tilde{v}(y))_k,
\end{equation}
and equations for $\tilde{v}_i$ are obtained by left-multiplying $(M\cdot M^\intercal)^{-1}M$ on both sides, the initial condition is $(M\cdot M^\intercal)^{-1}M v(y_{UV})$ where $v(y_{UV})$ is obtained in \eqnref{eq:initial_cond}. The RG equations for $\tilde{v}_i$ can reproduce the phase diagram in the previous section, but there are still too many terms. We further truncate the RG equations by ignoring the terms with coefficients much less than the leading one. For the displacement field $u_d\approx \SI{34}{\meV}$ with additional particle-hole nesting, we obtain the projected renormalization group equation as,
\begin{align*}
\frac{d\ivct}{dy} &=6 d_1 \ivct^2+2 d_1 i \ivct-2 d_1 \ivct s -d_1 f \ivct+d_1 f' \ivct-\frac{d_1 f f'}{7}-\frac{2 d_1 f i}{7}+\frac{2 d_1 f s}{7}+\frac{2 d_1 f' i}{7}\\
&-\frac{2 d_1 f' s}{7}+\frac{2 d_1 {i}^2}{7}-\frac{4 d_1 {i} s}{7}+\frac{2 d_1 s^2}{7} \\
 \frac{d\ivcs}{dy} &= 12 d_1 \ivcs^2+3 d_1 f' \ivcs-3 d_1 f \ivcs+2 d_1 \ivcs s-2 d_1 {i} \ivcs+\frac{9 d_1 f^2}{28}-\frac{9 d_1 f f'}{14}+\frac{3 d_1 f {i}}{7}-\frac{3 d_1 f s}{7}\\
 &+\frac{9 d_1 f'^2}{28}-\frac{3 d_1 f' {i}}{7}+\frac{3 d_1 f' s}{7}-\frac{2 d_1 {i} s}{7} \\
 \frac{ds}{dy} &=12 d_0 s^2+2 d_0 \ivcs s-3 d_0 \ivct s  -\frac{3 d_0 \ivcs \ivct}{8}+\frac{9 d_0 \ivct^2}{32}\\
 \frac{di}{dy} &= 12 d_0 {i}^2-2 d_0 {i} \ivcs+3 d_0 {i} \ivct-\frac{3 d_0 \ivcs \ivct}{8}+\frac{9 d_0 \ivct^2}{32} \\
 \frac{df}{dy} &= 6 d_0 f^2-2 d_0 f \ivcs-d_0 f \ivct+\frac{d_0 \ivcs^2}{4}+\frac{d_0 \ivcs \ivct}{4}-\frac{5 d_1 f f'}{28}-\frac{d_1 {i} s}{7} \\
 \frac{df'}{dy} &= 6 d_0 f'^2+2 d_0 f' \ivcs+d_0 f' \ivct+\frac{d_0 \ivcs^2}{4}+\frac{d_0 \ivcs \ivct}{4}+\frac{5 d_1 f f'}{28}+\frac{d_1 {i} s}{7}
\end{align*}
where $\ivct,\ivcs$ refer to spin-triplet IVC and spin-singlet IVC, $d_0=d\chi_{\text{pp}}^{K1,K'1}/dy,d_1=d\chi_{\text{ph}}^{K1,K6}/dy=d\chi_{\text{ph}}^{K1,K'6}/dy$ are the nesting parameters and other susceptibilities decay as fast as $1/y^{1/2}$ which can be ignored.
\begin{figure}[htbp]
    \centering
    \includegraphics[width=0.9\textwidth]{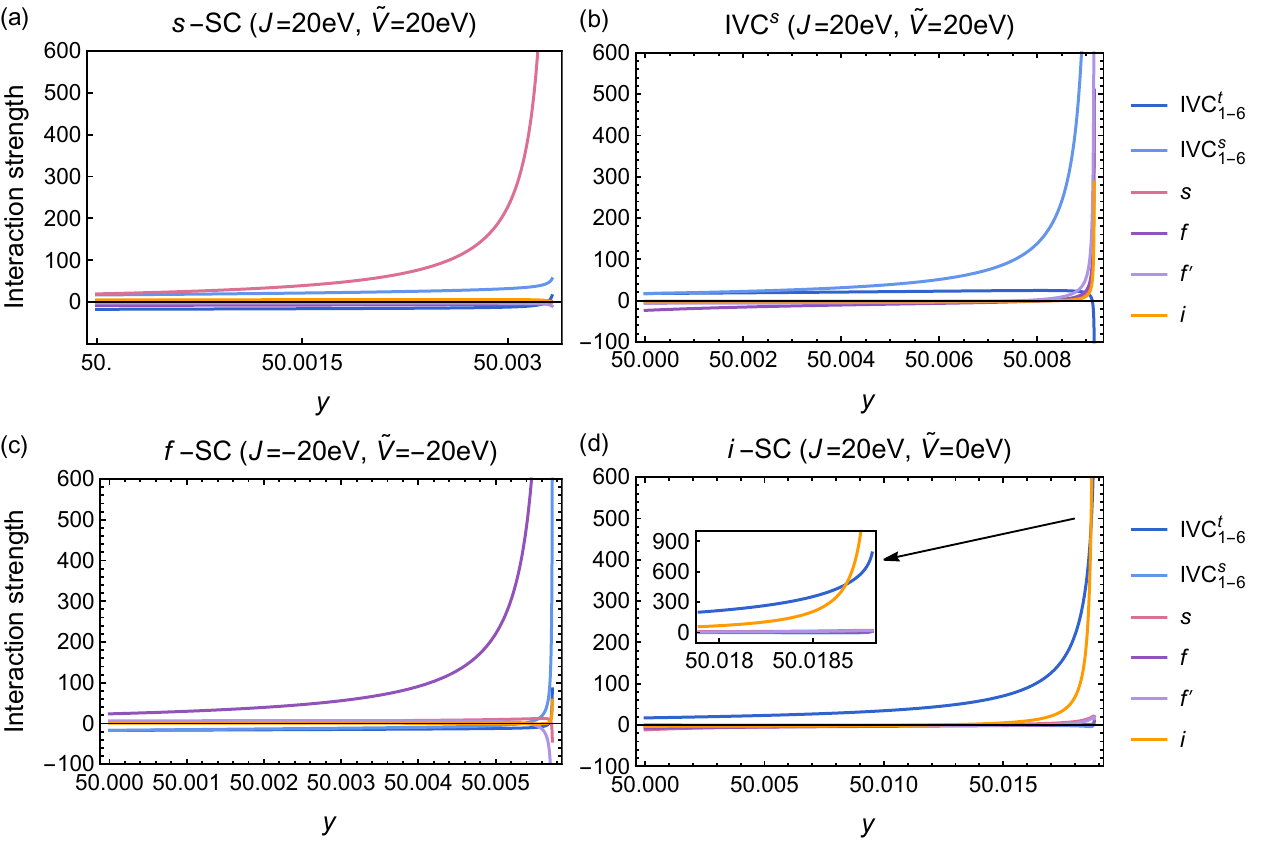}
    \caption{(a) - (d) show the RG flow of interaction vertices projected to the instability directions. (a) - (c) show the leading divergences are $s$-wave superconductivity, spin-singlet IVC and $f$-wave superconductivity respectively. (d) shows the competition between the spin-triplet IVC and $i$-wave superconductivity, the $i$-wave superconductivity finally diverges faster than any other instabilities.}
    \label{fig:rgdiag_34}
\end{figure}

We can see that both the IVC orders interplay with the different superconducting orders, in particular, the singlet IVC will enhance $s$ and $f'$-wave via the terms like $+2d_0 \ivcs s$ and $+2d_0 f' \ivcs$, and the triplet IVC will enhance $i$ by the term $3d_0 \ivct i$. This enhancement indeed happens in $i$-wave superconducting phase with $u_d\approx \SI{34}{\meV}$. As shown in \figref{fig:rgdiag_34} (d), the spin-triplet IVC ordering is the leading one at the high energy scale, under the RG flow, the energy scale gradually decreases and the $i$-wave superconductivity starts to dominate.

\subsection{Example of generating the RG equations}
\begin{figure}[htbp]
    \centering
    \includegraphics[width=0.7\textwidth]{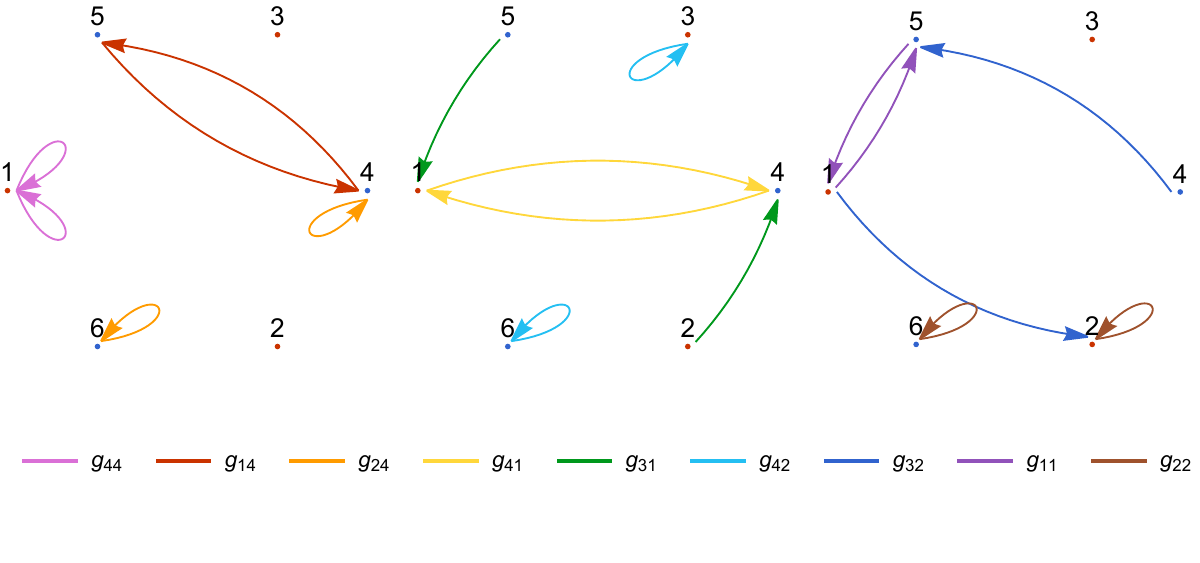}
    \caption{This figure shows 9 different four fermion interactions that satisfy momentum conservation. These are the representatives in each equivalence class. The naming is referring to \cite{isobe2018unconventional}.}
    \label{fig:vhs6diag}
\end{figure}

We demonstrate the way to generate the RG equations for the model with 3 VHSs in each valley, this model is relevant to the twisted bilayer graphene \cite{isobe2018unconventional}. There are 6 hot-spots in the first BZ, we label them as in \figref{fig:vhs6diag}. We first generate the tuples corresponding to the momentum conserved four fermion interaction vertices,
\begin{equation}
    \mathsf{IntAll}=\{(\taua{1},\taua{2},\taua{3},\taua{4})|\Ki{1}+\Ki{2}-\Ki{3}-\Ki{4}=0\}
\end{equation}
The VHSs are related to each other via group actions. If we denote the VHSs in the $K$ valley by $\alpha$ and those in the $K'$ valley by $\alpha+3$, the generators of the group actions can be represented as cycles,
\begin{equation}
    C_3: (132)(465),\quad M_x : (23)(56),\quad T: (14)(25)(36)
\end{equation}
These generators generate the permutation group with 12 elements, the elements in $\mathsf{IntAll}$ should relate to each other under these group actions. If two elements in $\mathsf{IntAll}$ are related to each other by,
\begin{equation}
    (\taua{x_1},\taua{x_2},\taua{x_3},\taua{x_4})=(g\cdot\taua{y_1},g\cdot\taua{y_2},g\cdot\taua{y_3},g\cdot\taua{y_4}), \  \exists g\in G,
\end{equation}
then they fall in the same equivalence class, the set $\mathsf{IntAll}$ then splits into several equivalence classes and we label the equivalence classes by their representatives $(\taua{1},\taua{2},\taua{3},\taua{4})^i$, where $i$ is the index for the equivalence class. On the contrary, all the interaction vertices can be organized as,
\begin{equation}
    \mathsf{IntSym}^i=\{(g\cdot\taua{1},g\cdot\taua{2},g\cdot\taua{3},g\cdot\taua{4})^i, \forall g \in G\}
\end{equation}
As shown in \figref{fig:vhs6diag}, it is interesting to draw the process of these representatives.

The $\mathsf{IntSym}$ gives all the momentum indices for the interaction vertices and organized by symmetry of the VHSs. Next, we need to add the flavor indices for the interaction vertices. For the sake of demonstration, we only consider additional spin degrees of freedom, and consider the four fermion interaction term as,
\begin{equation}
    H_\text{int}=g_i \sum_{\substack{(\taua{1},\taua{2},\taua{3},\taua{4})\\\in \mathsf{IntSym}^i}}\sum_{\sigma \sigma'}\psi_{\taua{4}\sigma}^\dagger \psi_{\taua{3}\sigma'}^\dagger \psi_{\taua{2}\sigma'}\psi_{\taua{1}\sigma}
\end{equation}
The four fermion interactions related under symmetry and fall in the same equivalence class $i$ will have the same strength $g_i$. Upon rewriting the complex fermions in terms of Majorana fermions via \eqnref{eq:majbasis}, the interaction vertex can be converted to totally antisymmetric rank-4 tensor, $V_{ABCD} \chi_A \chi_B \chi_C \chi_D \rightarrow V_{[ABCD]}$, where the bracket means antisymmetrization of the indices. It is convenient to define interaction vertex basis $u^i_{ABCD}$ as in the main text, therefore, the interaction term is,
\begin{equation}
    H_\text{int}=g_i u^i_{ABCD} \chi_A \chi_B \chi_C \chi_D.
\end{equation}
Then the one-loop correction for the interaction is given by \appref{app:RGEdetail}. The susceptibilities also have different equivalence classes under the symmetry of VHSs, for example, the susceptibility of fermion near VHSs $K 1$ and $K' 1$ are the same as $K 2$ and $K' 2$, and so on. As derived in \appref{app:RGEdetail}, the susceptibility can also be written as the rank-4 tensor. Therefore, the RG equation is obtained by contraction of the interaction tensor and the susceptibility tensor as shown in \figref{fig:oneloop}. All the tensors and the RG equations can be automatically generated by a computer and it is straightforward to generalize to the system with more VHSs.

The RG equations for the time reversal symmetric system with 3 VHSs in each valley are,

\begin{align*}
\dot{g}_{44}=&-g_{44}^2+d_{2-}\left(2 g_{11}^2+4 g_{22} g_{11}+2 g_{14}^2-4 g_{22}^2-4 g_{24}^2+g_{41}^2-2 g_{42}^2+g_{44}^2+4 g_{14} g_{24}+2 g_{41} g_{42}\right)  \\
 \dot{g}_{14}=&d_{1-}\left(-2 g_{14}^2+2 g_{24} g_{14}-2 g_{32}^2+2 g_{31} g_{32}\right)-2d_{3-} g_{14} g_{24}  +d_{2-}\left(g_{11}^2+2 g_{41} g_{11}+g_{14}^2+2 g_{14} g_{44}\right)  \\
 \dot{g}_{24}=&d_{1-}\left(g_{24}^2+g_{31}^2\right)+d_{3-}\left(-g_{14}^2-g_{24}^2\right)\\
 &+d_{2-}\left(-2 g_{22}^2+2 g_{11} g_{22}+2 g_{41} g_{22}-4 g_{42} g_{22}-2 g_{24}^2+2 g_{14} g_{24}+2 g_{11} g_{42}+2 g_{14} g_{44}-2 g_{24} g_{44}\right)   \\
 \dot{g}_{41}=& -4 g_{31} g_{32}-2 g_{41} g_{42}+d_{2-}\left(-2 g_{41}^2+2 g_{42} g_{41}+2 g_{44} g_{41}+4 g_{11} g_{14}\right) \\
 \dot{g}_{31}=&-2 g_{31} g_{32}-2 g_{41} g_{32}-2 g_{31} g_{42}+ d_{1-}\left(-4 g_{11} g_{31}+2 g_{22} g_{31}+2 g_{24} g_{31}+2 g_{11} g_{32}\right) \\
 \dot{g}_{42}=&-2 g_{31}^2-2 g_{32}^2-g_{41}^2-g_{42}^2 +d_{2-}\left(g_{42}^2-2 g_{44} g_{42}+4 g_{14} g_{22}+4 g_{11} g_{24}-8 g_{22} g_{24}+2 g_{41} g_{44}\right) \\
 \dot{g}_{32}=&-g_{31}^2-2 g_{41} g_{31}-g_{32}^2-2 g_{32} g_{42}+d_{1-}\left(2 g_{14} g_{31}-4 g_{14} g_{32}+2 g_{22} g_{32}+2 g_{24} g_{32}\right)  \\
 \dot{g}_{11}=&d_{1-} \left(-2 g_{11}^2+2 g_{22} g_{11}-2 g_{31}^2+2 g_{31} g_{32}\right) -2 d_{3-}g_{11} g_{22} +d_{2-}\left(2 g_{11} g_{14}+2 g_{41} g_{14}+2 g_{11} g_{44}\right)  \\
 \dot{g}_{22}=&d_{1-}\left(g_{22}^2+g_{32}^2\right)+d_{3-}\left(-g_{11}^2-g_{22}^2\right)   \\
 &+d_{2-}\left(2 g_{14} g_{22}-4 g_{24} g_{22}-2 g_{44} g_{22}+2 g_{11} g_{24}+2 g_{24} g_{41}+2 g_{14} g_{42}-4 g_{24} g_{42}+2 g_{11} g_{44}\right)
\end{align*}
where $\dot{g_i}=\frac{dg_i}{dy}$, where $y$ is the susceptibility of Cooper pair channel. Then the nesting parameter for the Cooper pair channel is $d\chi_\text{pp}^{K\alpha,K'\alpha}/dy=1$. $d_{1-}=d\chi_\text{pp}^{K 1,K' 2}/dy$ is the nesting parameter of the particle-hole channel between the VHS $K 1$ and $K' 2$ and other symmetry related pairs, $d_{3-}$ is the nesting parameter of the particle-particle channel between the VHS $K 1$ and $K 2$, and $d_{2-}$ is the nesting parameter of the particle-hole channel between the VHS $K 1$ and $K' 1$.

\end{document}